\documentclass[aps,prl,twocolumn,groupedaddress]{revtex4-1}

\usepackage{amsmath}

\newcommand{\be}{\begin{eqnarray}}
\newcommand{\ee}{\end{eqnarray}}
\newcommand{\ket}[1]{\ensuremath{\left| {#1} \right>}}

\newcommand{\expect}[1]{\ensuremath{\left< {#1} \right>}}

\newcommand{\Beplus}{\ensuremath{{\rm Be}^{+} \,}}
\newcommand{\Caplus}{\ensuremath{{\rm Ca}^{+} \,}}

\usepackage{latexsym}
\usepackage{graphicx}
\usepackage{multirow}
\usepackage{textcomp}

\usepackage{tikz}
\usetikzlibrary{arrows, arrows.meta, positioning, calc, shadows, decorations.markings}
\tikzset{>=Stealth}

\usepackage{hyperref}

\begin{document}
\title{Repeated multi-qubit readout and feedback with a mixed-species trapped-ion register}

\author{V.~Negnevitsky}
\thanks{These authors contributed equally to this work.}
\author{M.~Marinelli}
\thanks{These authors contributed equally to this work.}
\author{K.~Mehta}
\author{H-Y.~Lo}
\author{C.~Fl{\"u}hmann}
\author{J.~P.~Home }

\address{Institute for Quantum Electronics, ETH Z\"urich, Otto-Stern-Weg 1, 8093 Z\"urich, Switzerland}

\maketitle
\begin{bf}Quantum error correction will be essential for realizing the full potential of large-scale quantum information processing devices \cite{steane_error_1996, shor_scheme_1995}.
  Fundamental to its experimental realization is the repetitive detection of errors via projective measurements of quantum correlations among qubits, and correction using conditional feedback \cite{gottesman_theory_1998}.
  Performing these tasks repeatedly requires a system in which measurement and feedback decision times are short compared to qubit coherence times, where the measurement reproduces faithfully the desired projection, and for which the measurement process has no detrimental effect on the ability to perform further operations.
  Here we demonstrate up to 50 sequential measurements of correlations between two beryllium-ion qubits using a calcium ion ancilla, and implement feedback which allows us to stabilize two-qubit subspaces as well as Bell states.
  Multi-qubit mixed-species gates are used to transfer information from qubits to the ancilla, enabling quantum state detection with negligible crosstalk to the stored qubits.
  Heating of the ion motion during detection is mitigated using sympathetic re-cooling.
  A key element of the experimental system is a powerful classical control system, which features flexible in-sequence processing to implement feedback control.
  The methods employed here provide a number of essential ingredients for scaling trapped-ion quantum computing, and provide new opportunities for quantum state control and entanglement-enhanced quantum metrology \cite{leibfried_toward_2004}.
\end{bf}

Correction of errors arising from noise and imperfect gate operations represents a primary challenge in the development of quantum computers \cite{terhal_quantum_2015}.
Quantum error correction (QEC) involves encoding the states of one logical qubit onto codewords on multiple physical qubits.
Repeated measurement of multi-qubit correlations between a subset of the physical qubits in the code allows errors to be detected without disturbing the stored information.
In order to reverse the detrimental effect of errors, information regarding the measurement result must be processed in real time such that appropriate correction operations can be chosen and applied.
Such a feedback allows the quantum system to be stabilized throughout the extended durations required for computation \cite{steane_overhead_2003}.

QEC codes have now been demonstrated in a number of systems \cite{cory_experimental_1998, chiaverini_realization_2004, reed_realization_2012}.
Many early demonstrations relied on protocols that brought qubits out of the code-space for error syndrome measurement; for practical QEC on continuously encoded information, ideal von-Neumann projective measurements of multi-qubit operators are required \cite{gottesman_theory_1998}.
These have been realized with trapped ions \cite{barreiro_open-system_2011, schindler_experimental_2011, nigg_quantum_2014} and in solid state systems \cite{pfaff_demonstration_2013}, including experiments in which a single round of feedback was conditioned on the result of measurement \cite{riste_deterministic_2013}.
However, in order to perform useful repeated stabilizer readout and feedback as required for indefinite stabilization of quantum systems, additional stringent demands on the system must be satisfied: i) the measurement time must be short compared to relevant decoherence times of the system, ii) the measurement process should not adversely affect the stored quantum information, and, iii) the process of measurement should not impede the ability to perform subsequent measurements \cite{sun_tracking_2014}.
Ancilla measurement reset typically involves dissipative interaction between quantum systems and the classical world over extended periods of time, which places strong requirements on the suppression of cross-talk between these processes and the data qubits.
Although up to three consecutive measurements have been performed on a multi-qubit trapped ion register by hiding data qubits in internal states which do not interact with the resonant light used for detection \cite{monz_realization_2016}, it appears challenging to implement this with high fidelity in large-scale systems.
An alternative approach is offered by the use of two species, one for the ancilla and one for data qubits, which ensures a high degree of  spectral isolation \cite{schmidt_spectroscopy_2005, ballance_hybrid_2015, gaebler_high-fidelity_2016}.
This also provides the possibility to mitigate errors due to ion heating and transport using sympathetic cooling \cite{barrett_sympathetic_2003, home_complete_2009}.
Non-destructive sampling of quantum information from multiple qubits has application beyond quantum computing, particularly in the field of metrology, where it provides direct access to correlation functions of the system \cite{hume_high-fidelity_2007}.

In this Letter, we demonstrate the repeated readout of two-qubit correlations stored in the ground-state hyperfine structure of beryllium ions ($^9$Be$^+$) using a calcium ($^{40}$Ca$^+$) ion ancilla.
Single and multi-qubit operations are performed on the three co-trapped ions with pairs of Raman beams for \Beplus and a narrow-linewidth 729\,nm laser for \Caplus (see methods).
We achieve beryllium qubit coherence times long compared to all operations by encoding quantum information in a field-insensitive state of \Beplus \cite{langer_long-lived_2005}, while cross-talk from  measurement of the calcium ancilla is minimized by performing readout in a different species \cite{ballance_hybrid_2015, gaebler_high-fidelity_2016, bruzewicz_high-fidelity_2017}.
Ancilla recycling is performed by optical pumping of the calcium ion after each measurement, while a combination of electromagnetically-induced transparency (EIT) \cite{roos_experimental_2000, tan_multi-element_2015} and sideband cooling is used to re-initialize the motional ``quantum bus'' which is used for quantum gates between the \Beplus and \Caplus qubits.
These tools allow repeated measurements for more than 50 cycles.
In addition, we utilize a control system with a central-processing-unit to apply feedback operations in real-time conditioned on measurement results, by controlling both the \Beplus operations and the potentials applied to the ion trap.
Sequential measurements in different bases allow preparation of Bell states from arbitrary input states, with feedback allowing entanglement to be stabilized over extended periods.

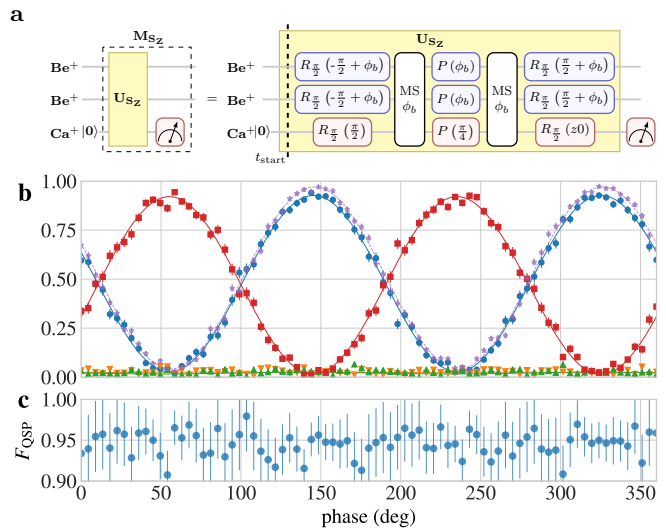
\begin{figure}[tb]
  \raggedright
  \textbf{a}

  \resizebox{1\linewidth}{!}{\def\minNodeH{0.6cm}
\def\minNodeHH{2cm}
\def\minNode2H{1.3cm}

\def\rowsp{0.7cm}
\def\colsp{0.9cm}

\def\labeloffs{4.5}
\def\linestart{4.1}

\begin{tikzpicture}[
  node distance=0.2cm,
  qnd_block_rect/.style={draw=black!30!yellow!100, thin, fill=yellow!30},
  qubit_line/.style={text=black, draw=black!20,very thick},
  comment_line/.style={text=black,dashed, draw=black,very thick},
  sqBe_coco_opt/.style={draw, thick, dashed, draw=black!20!blue!50, fill=blue!1, minimum height=\minNodeH},  
  sqBe_coco/.style={draw, thick, draw=black!30!blue!60, fill=blue!3, minimum height=\minNodeH},
  sqBe_co90/.style={draw, thick, draw=black!30!blue!60, fill=blue!3, rounded corners, minimum height=\minNodeH},
  feed_box/.style={draw=black!10!black!100, thin, dashed, minimum height=2.3*\minNodeH},
  qnd_block/.style={draw=black!30!yellow!100, thin, fill=yellow!30, minimum height=\minNodeHH},
  sqCa/.style={draw, thick, draw=black!50!red!60, fill=red!4, rounded corners, minimum height=\minNodeH, align=center},
  sqBeCa/.style={draw, thick, draw=black, fill=white, rounded corners, minimum height=\minNodeHH, align=center},
  measCa/.pic={
  	\draw [sqCa] (0,0) +(-0.3,-0.3) rectangle +(0.3,0.3);
  	\draw (0,-0.2) ++ (0:0.25) arc [start angle=0, end angle=180, radius=0.25];
  	\fill (0,-0.2) circle (1pt);
  	\draw [-{Latex[scale=0.7]}] (0,-0.2) -- ++(70:0.45);
  },
  measBe/.pic={
	\draw [sqBe_co90] (0,0) +(-0.3,-0.3) rectangle +(0.3,0.3);
	\draw (0,-0.2) ++ (0:0.25) arc [start angle=0, end angle=180, radius=0.25];
	\fill (0,-0.2) circle (1pt);
	\draw [-{Latex[scale=0.7]}] (0,-0.2) -- ++(70:0.45);
  },
  ]
  \footnotesize

	\node at ($-\labeloffs*(\colsp,0)+(0,0)$) (be_label1){$\rm \bf Be^+$};
	\node at ($-\labeloffs*(\colsp,0)+(0,-\rowsp)$) (be_label1){$\rm \bf Be^+$}; 
	\node at ($-\labeloffs*(\colsp,0)+2*(0,-\rowsp)$) (ca_label){$\rm \bf Ca^+$};
	\node at ($-3.9*(\colsp,0)+2*(0,-\rowsp)$) (ca_label){$\bf |0\rangle$};
	\draw[qubit_line] ($-\linestart*(\colsp,0)$) -- +($2.7*(\colsp,0)$);
	\draw[qubit_line] ($-\linestart*(\colsp,0)+(0,-\rowsp)$) -- +($2.7*(\colsp,0)$);
	\draw[qubit_line] ($-3.7*(\colsp,0)+2*(0,-\rowsp)$) -- +($1.5*(\colsp,0)$);

	\node[qnd_block] at ($-3*(\colsp,0)-1*(0,\rowsp)$)  {$\bf U_{S_Z}$};
	\pic at ($-2*(\colsp,0)-2*(0,\rowsp)$) {measCa};

	\draw[feed_box] ($-3.6*(\colsp,0)+0.5*(0,\rowsp)+0.1*(0,\rowsp)$) rectangle ($-1.5*(\colsp,0)-2.6*(0,\rowsp)$);
	\node at ($-2.5*(\colsp,0)+0.9*(0,\rowsp)$) {$\bf M_{S_Z}$};

	\node at ($-1*(\colsp,0)-1.*(0,\rowsp)$) {$=$};


  \draw[qnd_block_rect] ($0.6*(\colsp,0)+0.6*(0,\rowsp)+0.5*(0,\rowsp)$) rectangle ($8.7*(\colsp,0)-2.6*(0,\rowsp)$);
  \node at ($4.2*(\colsp,0)+0.8*(0,\rowsp)$) {$\bf U_{S_Z}$};
  \node at ($-2.8*(0,\rowsp)+0.4*(\colsp,0)$) {$t_{\rm start}$};
  \draw[comment_line] ($0.8*(\colsp,0)+1.3*(0,\rowsp)$) -- +($-4.1*(0,\rowsp)$);

  \draw[qubit_line] ($0.2*(\colsp,0)$) -- +($9*(\colsp,0)$);
  \draw[qubit_line] ($0.2*(\colsp,0)+(0,-\rowsp)$) -- +($9*(\colsp,0)$);
  \draw[qubit_line] ($0.4*(\colsp,0)+2*(0,-\rowsp)$) -- +($9.0*(\colsp,0)$);
  
  
  \node at ($-0.3*(\colsp,0)+(0,0)$) (be_label1){$\rm \bf Be^+$};
  \node at ($-0.3*(\colsp,0)+(0,-\rowsp)$) (be_label1){$\rm \bf Be^+$};
  \node at ($-0.3*(\colsp,0)+2*(0,-\rowsp)$) (ca_label){$\rm \bf Ca^+$};
  \node at ($0.2*(\colsp,0)+2*(0,-\rowsp)$) (ca_label){$\bf |0\rangle$};  
%
%
%
  \node[sqCa] at ($2.1*(\colsp,0)+2*(0,-\rowsp)$) {$R_{\frac{\pi}{2}}\left( \frac{\pi}{2}\right)$};
  \node[sqCa] at ($4.8*(\colsp,0)+2*(0,-\rowsp)$) {$P\left(\frac{\pi}{4}\right)$};
  \node[sqCa] at ($7.4*(\colsp,0)+2*(0,-\rowsp)$) {$R_{\frac{\pi}{2}}\left(z 0\right)$};
  \pic at ($9.2*(\colsp,0)-2*(0,\rowsp)$) {measCa};
    
  \node[sqBe_co90] at ($2.1*(\colsp,0)+0.0*(0,-\rowsp)$) {$R_{\frac{\pi}{2}}\left(\text{-}\frac{\pi}{2}+\phi_b\right)$};
  \node[sqBe_co90] at ($2.1*(\colsp,0)+1.0*(0,-\rowsp)$) {$R_{\frac{\pi}{2}}\left( \text{-}\frac{\pi}{2}+\phi_b\right)$};
  \node[sqBe_co90] at ($4.8*(\colsp,0)+0.0*(0,-\rowsp)$) {$P\left(\phi_b\right)$};
  \node[sqBe_co90] at ($4.8*(\colsp,0)+1.0*(0,-\rowsp)$) {$P\left(\phi_b\right)$};
  \node[sqBe_co90] at ($7.5*(\colsp,0)+0.0*(0,-\rowsp)$) {$R_{\frac{\pi}{2}}\left( \frac{\pi}{2}+\phi_b\right)$};
  \node[sqBe_co90] at ($7.5*(\colsp,0)+1.0*(0,-\rowsp)$) {$R_{\frac{\pi}{2}}\left( \frac{\pi}{2}+\phi_b\right)$};

  \node[sqBeCa] at ($3.7*(\colsp,0)+(0,-\rowsp)$) (ms1){MS \\ $\phi_b$};
  \node[sqBeCa] at ($5.9*(\colsp,0)+(0,-\rowsp)$) (ms2){MS \\ $\phi_b$};
 

\end{tikzpicture}}
  \includegraphics[width = \columnwidth]{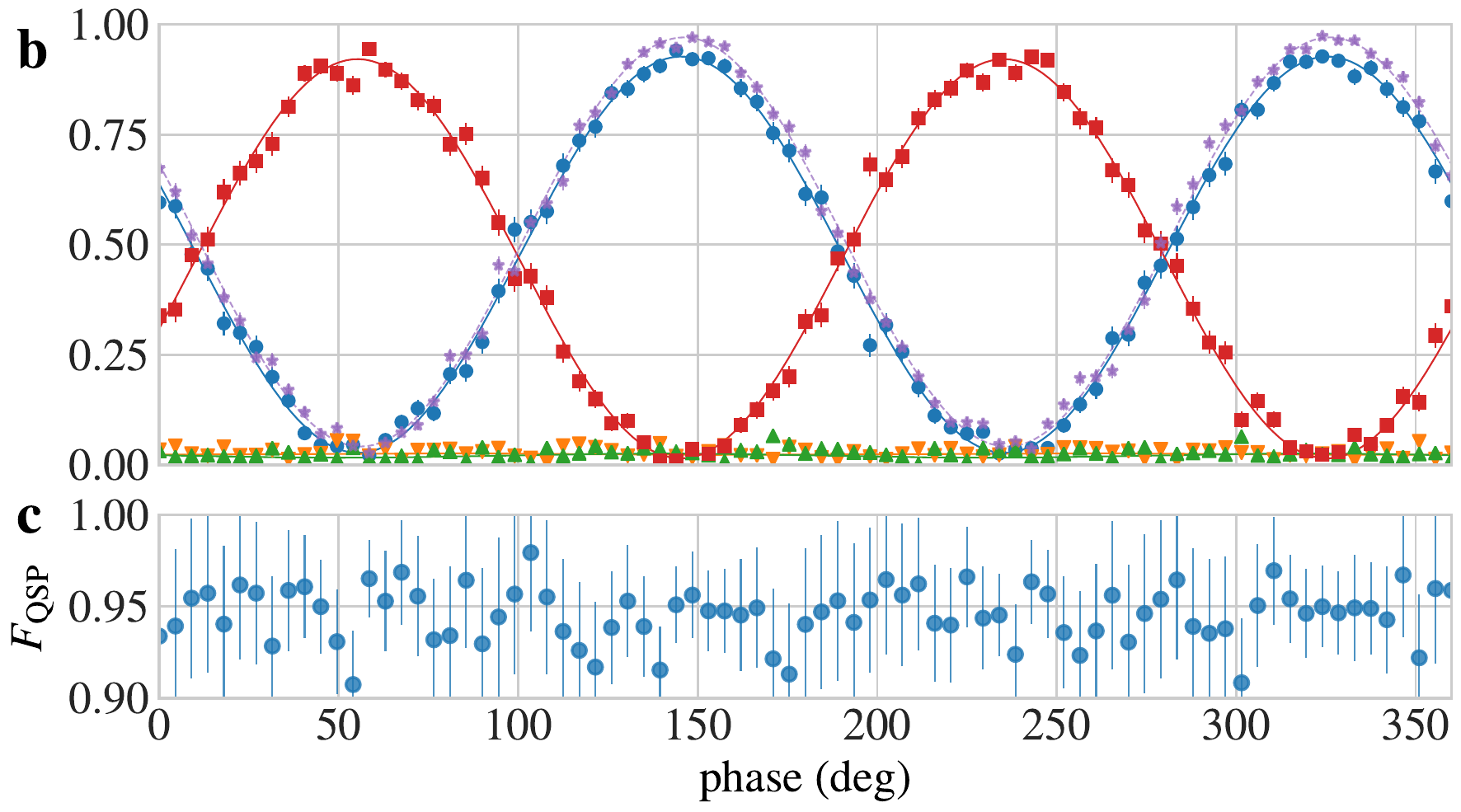}
  \caption{$S_Z$ parity measurement.
    \textbf{a.} The parity measurement $M_{S_Z}$ involves a unitary $U_{S_Z}$, which entangles the \Beplus parity eigenspaces with the \Caplus state, followed by a projective measurement of the latter.
    The decomposition of $U_{S_Z}$ is shown on the right, including two M\o lmer-S\o rensen gates, $R_{\pi/2}(\phi)$ basis rotations and inversion operations $P(\phi) \equiv \cos(\phi) X + \sin(\phi) Y$.
    The dashed vertical line indicates the reference time for the local phase accumulation, described further in the Methods section.
    The measurement basis can be rotated to $S_X$ using additional $R_{\pi/2}$ pulses (not shown).
    \textbf{b.} Data from the calcium detection of a single round of parity measurement combined with the direct detection of beryllium after the measurement as a function of the phase $\theta$ for the 2-ion \Beplus input state
    $\cos(2\theta)\ket{-} + \sin(2\theta)\ket{+}$, where $\ket{+} = \ket{00} + e^{i\phi(\theta)}\ket{11}$, $E_Z = +1$ and $\ket{-} = \ket{01} + \ket{10}$, $E_Z=-1$. $\theta$ is offset from the parameter $\phi_p$ in the main text due to an uncompensated Stark shift in the state preparation.
    Blue circles (red squares) show the probability of observing both that the \Beplus is in the $E_Z=+1$ ($E_Z=-1$) subspace and the \Caplus is measured to be in $\ket{1}$ ($\ket{0}$). 
    Triangles indicate the populations for which the calcium result is anti-correlated with respect to the ideal case.
    Violet stars show the parity of the input state, measured using direct beryllium detection in a separate experiment in which $M_{S_Z}$ was not implemented.
    Each point is the result of 300 experiments and error-bars are obtained assuming quantum projection noise.
    \textbf{c.} The corresponding quantum state fidelity indicating the mean conditional probability that the parity subspace in the beryllium detection is correct given the value obtained from $M_{S_Z}$.}
  \label{fig:qnd_parity}
\end{figure}

We measure the commuting stabilizer operators $S_Z = Z_1\otimes Z_2$ and $S_X =  X_1\otimes X_2$ (indices refer to the qubit number, while here and elsewhere $X,Y,Z$ are the Pauli operators) using the circuit shown in Fig.\,\ref{fig:qnd_parity}a \cite{barreiro_open-system_2011}.
Eigenstates of these operators with eigenvalue $E_{Z/X}$ = +1($-1$) have positive (negative) parity.
To map the value of the parity onto the calcium measurement basis, we apply the multi-qubit operation $U_{S_Z}$ to the beryllium qubits with the calcium ion initialized in $\ket{0}$.
A subsequent projective measurement of calcium using state-dependent fluorescence then completes the stabilizer readout.
For an ideal implementation $M_{S_Z}$, this projects \Beplus into a $+1 (-1)$ eigenstate of $S_Z$ correlated with the \Caplus ion detected in $\ket{1}$ ($\ket{0}$).
At the core of $U_{S_Z}$ is a diagonal operation in the computational basis which can be written as $\exp{(i \pi Z_{\rm Ca} \otimes S_Z/4)}$, and which is implemented using two multi-species three-qubit M\o lmer-S\o rensen (MS) gates \cite{sorensen_entanglement_2000} performed using the in-phase motional mode plus additional single-qubit rotations (see Fig.\,\ref{fig:qnd_parity}a ) \cite{barreiro_open-system_2011}.
To map the phase shift of the calcium ion to the readout basis, we embed this operation between $\pi/2$ rotations (defined as  $R_{\pi/2}(\phi) \equiv \exp(i \pi P(\phi)/2)$ with $P(\phi) = \cos(\phi) X + \sin(\phi) Y$), which differ in phase by $\pi/2$.
This can be interpreted as a Ramsey interference experiment.
The resulting unitary is  $U_{S_Z} = R_{\pi/2}(\phi+\pi/2).\exp{(i \pi Z_{\rm Ca}S_Z)}.R_{\pi/2}(\phi)$.

Readout of $S_X$ is performed by a unitary $U_{S_X}$ which is formed by embedding $U_{S_Z}$ between $R_{\pi/2}$ basis rotations of both beryllium qubits about the $Y$ axis.
These are performed using Raman transitions driven by two co-propagating laser beams.
While at first sight it might appear that a reduced pulse sequence could be used to read out $S_X$, we operate under the experimental constraint that the parity readout must perform the same operation at any point in the sequence when it is applied.
The MS gate between the ground-state hyperfine qubits of the two \Beplus ions and the optical \Caplus qubit uses a pair of beams for the \Beplus hyperfine qubit-motion coupling which propagate to the ions with a relative angle of 90 degrees, and which are not phase-stable over many rounds of measurement.
The gate produces the operation ${\rm MS} \equiv \exp{(i \pi \Pi_{\phi_b}^2 /8)}$ with $\Pi_{\phi_b} = X_{\rm Ca} + P_1(\phi_b) + P_2(\phi_b)$.
$\phi_b$ is directly proportional to the Raman beam difference phase.
Since its drift is small on the 160 $\mu$s timescale of a single application of the unitary, we can mitigate it by using the same Raman beam pair to perform $\pi/2$ qubit rotations before ${\rm MS}_1$ and after ${\rm MS}_2$.
This makes the unitary diagonal in the computational basis, and thus ensures insensitivity to $\phi_b$  \cite{lee_phase_2005, gaebler_high-fidelity_2016}.
This choice also allows us to simplify pulse sequence control (see Methods).

To test the performance of a single round of the parity measurement $M_{S_Z}$, we input states with a range of parities, and compare the \Caplus detection results to the results of a subsequent state detection performed directly on the \Beplus ions.
To prepare input states, we create the Bell state $\left(\ket{00} - i \ket{11} \right)/\sqrt{2}$ using an MS gate acting only on the two \Beplus ions, and then apply an $R_{\pi/2}(\phi_p)$ pulse to both ions with phase $\phi_p$.
For $\phi_p = 3 \pi/4$, the resulting state has  $E_Z = -1$, while for $\phi_p = \pi/4$, $E_Z = +1$.
Results for the detection of both \Caplus and \Beplus measurements are shown in Fig.\,\ref{fig:qnd_parity}b.
For comparison, we also plot the parity obtained from measuring the input state of the \Beplus ions in a separate reference experiment.
We can compare the probability distributions estimated from the \Beplus measurements with and without the $M_{S_Z}$ parity measurement using a classical fidelity, for which the average value across all input states is $\overline{F}_{\rm ND} = 98.4(2)\%$.
A more important quality measure for QEC is the conditional probability for a given detection result to project the system into the correct subspace.
As shown in  Fig.\,\ref{fig:qnd_parity}c, this is largely independent of input state, with a mean value of $\overline{F}_{\rm QSP} = 94.5(3)\%$.
These results are consistent with the quality of operations in our system.
We find that we can create Bell states of two beryllium ions in the 3-ion chain with fidelities of up to $97.8(4)\%$ and that the fidelity for creating maximally entangled 3-qubit Greenberger-Horne-Zeilinger states including also the calcium ion up to  93.8(5)\%.

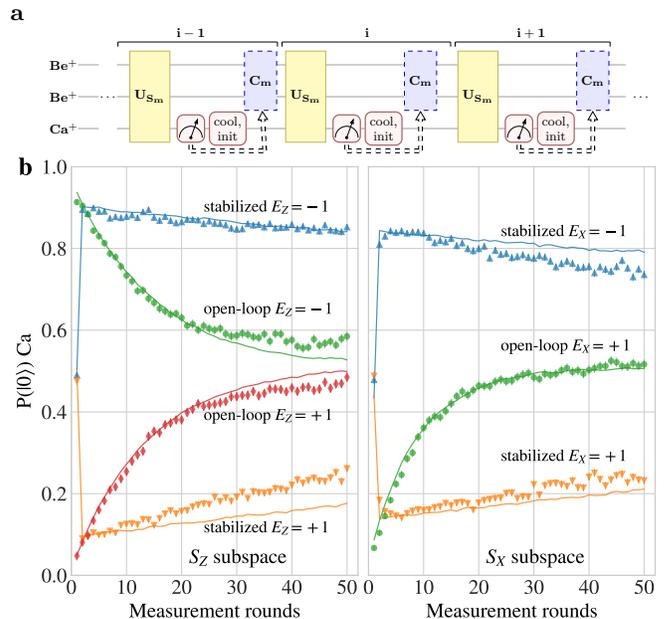
\begin{figure}[tb]
  \textbf{a}
  \raggedright
  \resizebox{1\linewidth}{!}{\def\minNodeH{0.6cm}
\def\minNodeHH{2cm}
\def\minNode2H{1.3cm}

\def\rowsp{0.7cm}
\def\colsp{0.9cm}
\def\labeloffs{4.1}
\def\linestart{3.75}
\begin{tikzpicture}[
  node distance=0.2cm,
  qnd_block/.style={draw=black!30!yellow!100, thin, fill=yellow!30, minimum height=\minNodeHH},
  feedback_block/.style={draw=black!10!blue!100, thin, dashed, fill=blue!10, minimum height=\minNode2H},
  qubit_line/.style={text=black, draw=black!20,very thick},
  sqBe_coco_opt/.style={draw, thick, dashed, draw=black!20!blue!50, fill=blue!1, minimum height=\minNodeH},  
  sqBe_coco/.style={draw, thick, draw=black!30!blue!60, fill=blue!3, minimum height=\minNodeH},
  sqBe_co90/.style={draw, thick, draw=black!30!blue!60, fill=blue!3, rounded corners, minimum height=\minNodeH},
  sqCa/.style={draw, thick, draw=black!50!red!60, fill=red!4, rounded corners, minimum height=\minNodeH, align=center},
  sqCaMeas/.style={draw, thick, draw=black!50!red!60, fill=red!4, minimum height=\minNodeH},  
  measCa/.pic={
    \draw [sqCa] (0,0) +(-0.3,-0.3) rectangle +(0.3,0.3);
    \draw (0,-0.2) ++ (0:0.25) arc [start angle=0, end angle=180, radius=0.25];
    \fill (0,-0.2) circle (1pt);
    \draw [-{Latex[scale=0.7]}] (0,-0.2) -- ++(70:0.45);
  },
	vecArrow/.style ={thick, dashed, decoration={markings,mark=at position
	1 with {\arrow[semithick]{open triangle 60}}},
	double distance=1.4pt, shorten >= 5.5pt,
	preaction = {decorate},
	postaction = {draw,line width=1.4pt, white,shorten >= 4.5pt}
	}
  ]
  \footnotesize
  

  \node at ($-\labeloffs*(\colsp,0)+(0,0)$) (be_label1){$\rm \bf Be^+$};
  \node at ($-\labeloffs*(\colsp,0)+(0,-\rowsp)$) (be_label1){$\rm \bf Be^+$};
  \node at ($-\labeloffs*(\colsp,0)+2*(0,-\rowsp)$) (ca_label){$\rm \bf Ca^+$};
  \draw[qubit_line] ($-\linestart*(\colsp,0)$) -- +($0.5*(\colsp,0)$);
  \draw[qubit_line] ($-\linestart*(\colsp,0)+(0,-\rowsp)$) -- +($0.5*(\colsp,0)$);
  \draw[qubit_line] ($-\linestart*(\colsp,0)+2*(0,-\rowsp)$) -- +($0.5*(\colsp,0)$);
  
  \node at  ($-3*(\colsp,0)-1*(0,\rowsp)$) (leftellipsis) {\ldots};
  \draw[qubit_line] ($-2.8*(\colsp,0)$) -- +($12.4*(\colsp,0)$);
  \draw[qubit_line] ($-2.8*(\colsp,0)+(0,-\rowsp)$) -- +($12.4*(\colsp,0)$);
  \draw[qubit_line] ($-2.8*(\colsp,0)+2*(0,-\rowsp)$) -- +($12.4*(\colsp,0)$);

  	\node[qnd_block](Um) at ($-2*(\colsp,0)-1*(0,\rowsp)$)  {$\bf U_{S_m}$};
  	\pic at ($-1.0*(\colsp,0)-2*(0,\rowsp)$) {measCa};
  	\node[sqCa] at ($-0.1*(\colsp,0)-2*(0,\rowsp)$) {cool,\\init};
  	\node[feedback_block](feed) at ($0.7*(\colsp,0)-0.5*(0,\rowsp)$)  {$\rm \bf C_m$};
  	\node[inner sep=0,minimum size=0] at ($-1*(\colsp,0)-2.5*(0,\rowsp)$) (meas) {}; 
  	\node[inner sep=0,minimum size=0] at ($-1*(\colsp,0)-2.7*(0,\rowsp)$) (int) {}; 
  	\node[inner sep=0,minimum size=0] at ($0.8*(\colsp,0)-2.7*(0,\rowsp)$) (int2) {}; 
  	\draw[vecArrow] (meas) |- (int) to (int2) to(feed);
  	\draw (-2.5,+0.5) -- (1,0.5);
  	\draw (-2.5,+0.5) -- (-2.5,0.4);
  	\draw (1,+0.5) -- (1,0.4);
  	\node at ($-1.0*(\colsp,0)+1*(0,\rowsp)$) {$\bf i-1$};

	\node[qnd_block](Um) at ($1.8*(\colsp,0)-1*(0,\rowsp)$)  {$ \bf U_{S_m}$};
	\pic at ($2.8*(\colsp,0)-2*(0,\rowsp)$) {measCa};
	\node[sqCa] at ($3.7*(\colsp,0)-2*(0,\rowsp)$) {cool,\\init};
	\node[feedback_block](feed2) at ($4.6*(\colsp,0)-0.5*(0,\rowsp)$)  {$\rm \bf C_m$};
	\node[inner sep=0,minimum size=0] at ($2.8*(\colsp,0)-2.5*(0,\rowsp)$) (meas) {}; 
	\node[inner sep=0,minimum size=0] at ($2.8*(\colsp,0)-2.7*(0,\rowsp)$) (int) {}; 
	\node[inner sep=0,minimum size=0] at ($4.6*(\colsp,0)-2.7*(0,\rowsp)$) (int2) {}; 
	\draw[vecArrow] (meas) |- (int) to (int2) to(feed2);
	\draw (1.1,+0.5) -- (4.8,0.5);
	\draw (1.1,+0.5) -- (1.1,0.4);
	\draw (4.8,+0.5) -- (4.8,0.4);
	\node at ($3.3*(\colsp,0)+1*(0,\rowsp)$) {$\bf i$};

	\node[qnd_block](Um) at ($6*(\colsp,0)-1*(0,\rowsp)$)  {$\bf U_{S_m}$};
	\pic at ($7*(\colsp,0)-2*(0,\rowsp)$) {measCa};
	\node[sqCa] at ($7.9*(\colsp,0)-2*(0,\rowsp)$) {cool,\\init};
	\node[feedback_block](feed3) at ($8.8*(\colsp,0)-0.5*(0,\rowsp)$)  {$\rm \bf C_m$};
	\node[inner sep=0,minimum size=0] at ($7*(\colsp,0)-2.5*(0,\rowsp)$) (meas) {}; 
	\node[inner sep=0,minimum size=0] at ($7*(\colsp,0)-2.7*(0,\rowsp)$) (int) {}; 
	\node[inner sep=0,minimum size=0] at ($8.8*(\colsp,0)-2.7*(0,\rowsp)$) (int2) {}; 
	\draw[vecArrow] (meas) |- (int) to (int2) to(feed3);
	\draw (4.9,+0.5) -- (8.3,0.5);
	\draw (8.3,+0.5) -- (8.3,0.4);
	\draw (4.9,+0.5) -- (4.9,0.4);
	\node at ($7.3*(\colsp,0)+1*(0,\rowsp)$) {$\bf i+1$};
	 
	\node at  ($10*(\colsp,0)-1*(0,\rowsp)$) (leftellipsis) {\ldots};

\end{tikzpicture}}
  \raggedright  
  \includegraphics[width = \columnwidth]{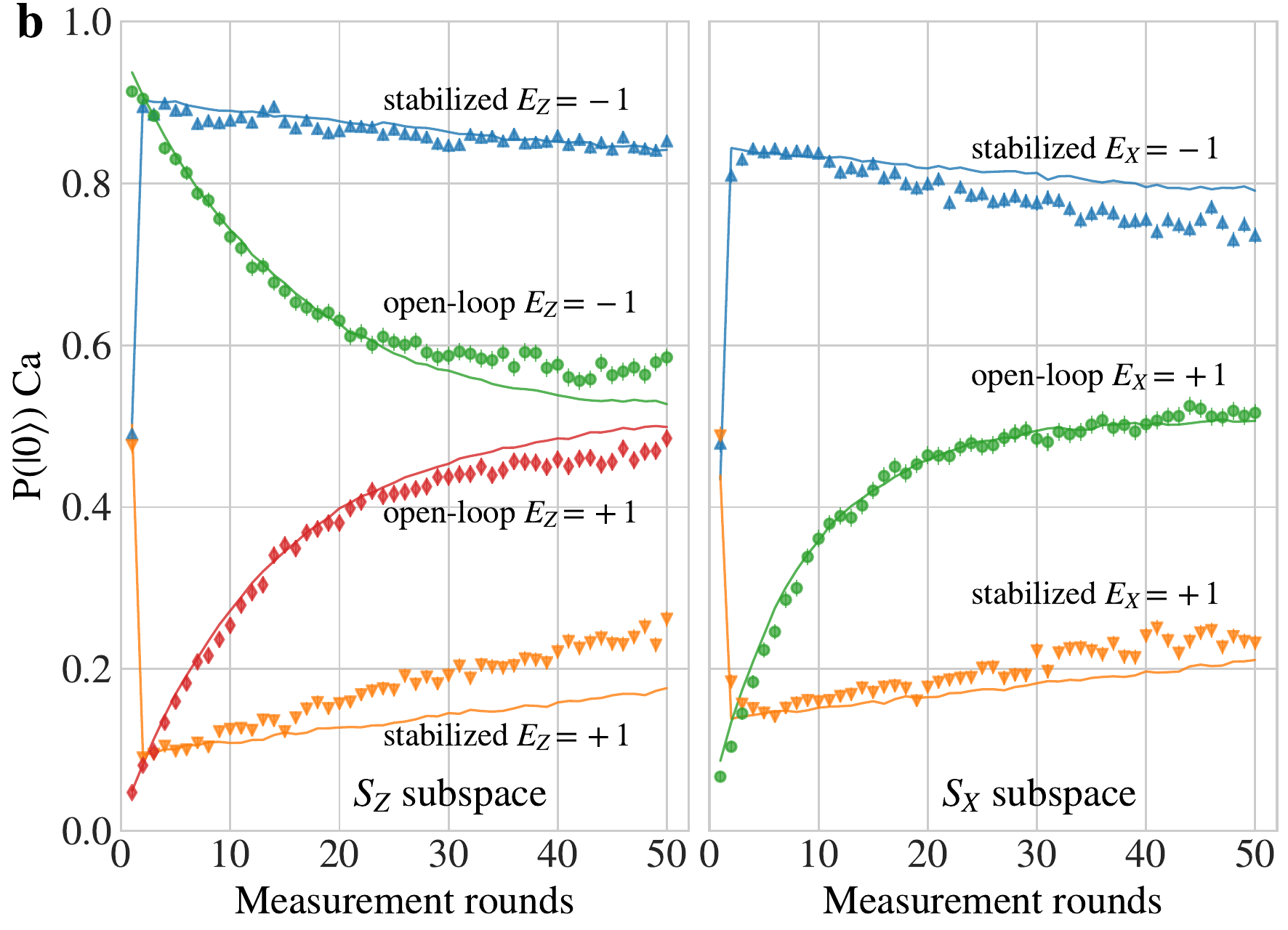}
  \caption{Repeated parity measurement and parity subspace stabilization.
    \textbf{a.} Illustration of the sequence used for stabilization of a subspace, showing the use of cooling and feedback.
    Depending on the choice of stabilization, the unitary is $U_{S_Z}$ or $U_{S_X}$, with the corresponding conditional feedback operations $C_Z$ or $C_X$ respectively.
    Recooling is performed using the \Caplus ancilla.
    \textbf{b.} \Caplus detection outcome probabilities as a function of repeated parity readouts for the $S_Z$ subspace.
    The circles (diamonds) show the outcomes as a function of repeated measurement rounds for the $S_Z$ without feedback when initially preparing an $E_Z=-1$ ($E_Z=+1$) \Beplus input state.
    The upward- (downward)-pointing triangles show the results with feedback conditioned on the \Caplus results, stabilizing the two $E_Z = +1$ and $E_Z = -1$ subspaces.
    \textbf{c.} Similar data for stabilization of the $S_X$ subspace.
    In this case circles show the outcomes without feedback after preparing an $E_X=+1$ input state.
    Each data point is the result of 2000 measurements, resulting in statistical uncertainties which are smaller than markers.
    Solid lines in (b) and (c) are produced using a simple Monte-Carlo density matrix simulation using only one free parameter (see Methods for discussion about the model and the discrepancies with measured data).
  }
  \label{fig:subspace_stab}
\end{figure}

Prior to a second round of measurement, we re-initialize the \Caplus qubit and the axial motional modes which affect the performance of the multi-qubit gates.
Fluorescence detection of \Caplus scatters photons, which induces considerable heating of both the in-phase and ``Egyptian'' modes of motion.
Due to the symmetry of our ion chain it does not heat the out-of-phase mode of motion of the two \Beplus ions (see Methods).
However the modes which are heated can also be sympathetically cooled by \Caplus, which we perform using  EIT cooling followed by a few cycles of pulsed sideband cooling \cite{monroe_resolved-sideband_1995}.
We then optically pump the \Caplus ion to $\ket{0}$, allowing a subsequent round of parity readout.
Fig.\,\ref{fig:subspace_stab} shows the \Caplus readouts for up to 50 sequential parity measurements performed on a pair of ions initially prepared in the $E_Z=\pm 1$ subspace.

By introducing feedback correction conditioned on the $M_{S_Z}$ or $M_{S_X}$ outcomes, we demonstrate the ability to stabilize a parity subspace.
The correction operation for $S_Z$ is $C_Z = -I_1\otimes X_2$, while for the $S_X$ eigenspaces we implement $C_X = -I_1 \otimes Z_2$.
In each case this requires a differential rotation of the \Beplus qubits.
This is implemented by changing the potentials of the trap to shift the three-ion chain in the \Beplus Raman control laser beam such that one ion experiences twice the Rabi frequency of the other.
At that position, we use a pulse duration which performs an $X$ operation on one qubit while performing a $-I$ operation on the other.
These pulses also produce undesired but stable AC Stark shifts, which we manage by updating the Raman laser differential phase for subsequent coherent operations.
$I\otimes Z$ is obtained from $I\otimes X$ using common $R_{\pi/2}$ rotations about the $Y$ axis of the qubits before and after the addressed operation to switch the operation basis.
Corrections are applied on each occasion that the \Caplus measurement indicates the undesired subspace.
Results from measurements in which the feedback stabilization is applied are shown in Fig.\,\ref{fig:subspace_stab}b.
The initial state in each of these runs is produced using a $\pi/2$ rotation applied to both beryllium ions after initialization of one of the computational basis states, which is an equal superposition of eigenstates with $E_Z = \pm 1$.
We see that the data exhibits decay constants of around 8\% per measurement round, which is consistent with the quality of measurements in regular operation of the experimental system.

To obtain a figure of merit for the improvement offered by the stabilization, we compare the decay rate of exponential fits to the open-loop data with that obtained in closed-loop by a linear fit to all data apart from the first point.
In a comparison of these rates, we see a significant improvement of order $\gamma_{\rm open}/\gamma_{\rm closed} \sim 20$.
The decay observed for the closed-loop data is at a level similar to what we can ascribe to a simulation including an independently measured level of leakage from the qubit subspace of the beryllium ions into neighbouring hyperfine levels, as produced by spontaneous photon scattering (solid lines in Fig.\,\ref{fig:subspace_stab}b, see Methods).
This does not fully account for the decay, although the addition of a gradual change in the parity readout fidelity of 0.06\% per measurement round does produce much better agreement.
This level is within what might be produced by effects such as thermal cycling of acousto-optic modulators which control the laser beam intensities.
We note that leakage is expected to be similar to the rate of error due to Raman scattering within the qubit subspace.
Although the gates in our experiments are not currently limited by this error source, once other sources of error are eliminated \cite{ballance_high-fidelity_2016, gaebler_high-fidelity_2016} this will become an important factor, which is not easily corrected by standard techniques in QEC, and may favour the use of ions without hyperfine structure  \cite{terhal_quantum_2015, brown_comparing_2018}.

\begin{figure} [htb]
  \raggedright
  \textbf{a}
  \resizebox{1\linewidth}{!}{\def\minNodeH{0.6cm}
\def\minNodeHH{2cm}
\def\minNode2H{1.3cm}

\def\rowsp{0.7cm}
\def\colsp{0.9cm}

\def\labeloffs{5.1}
\def\linestart{4.75}

\begin{tikzpicture}[
  node distance=0.2cm,
  qnd_block/.style={draw=black!30!yellow!100, thin, fill=yellow!30, minimum height=\minNodeHH},
  feedback_block/.style={draw=black!10!blue!100, thin, dashed, fill=blue!10, minimum height=\minNode2H},
  feedback_block_global/.style={draw=black!10!blue!100, thin, dashed, fill=blue!10, minimum height=2*\minNodeH},
  feed_box/.style={draw=black!10!black!100, thin, dashed, minimum height=2.3*\minNodeH},
  qubit_line/.style={text=black, draw=black!20,very thick},
  sqBe_coco_opt/.style={draw, thick, dashed, draw=black!20!blue!50, fill=blue!1, minimum height=\minNodeH},  
  sqBe_coco/.style={draw, thick, draw=black!30!blue!60, fill=blue!3, minimum height=\minNodeH},
  sqBe_co90/.style={draw, thick, draw=black!30!blue!60, fill=blue!3, rounded corners, minimum height=\minNodeH},
  sqCa/.style={draw, thick, draw=black!50!red!60, fill=red!4, rounded corners, minimum height=\minNodeH, align=center},
  sqCaMeas/.style={draw, thick, draw=black!50!red!60, fill=red!4, minimum height=\minNodeH},  
  measCa/.pic={
    \draw [sqCa] (0,0) +(-0.3,-0.3) rectangle +(0.3,0.3);
    \draw (0,-0.2) ++ (0:0.25) arc [start angle=0, end angle=180, radius=0.25];
    \fill (0,-0.2) circle (1pt);
    \draw [-{Latex[scale=0.7]}] (0,-0.2) -- ++(70:0.45);
  },
	vecArrow/.style ={thick, dashed, decoration={markings,mark=at position
	1 with {\arrow[semithick]{open triangle 60}}},
	double distance=1.4pt, shorten >= 5.5pt,
	preaction = {decorate},
	postaction = {draw,line width=1.4pt, white,shorten >= 4.5pt}
	}
  ]
  \footnotesize
  
  
\node at ($-4.4*(\colsp,0)+(0,0)$) (be_label1){$\rm \bf Be^+$};
\node at ($-4.4*(\colsp,0)+(0,-\rowsp)$) (be_label1){$\rm \bf Be^+$};
\node at ($-4.4*(\colsp,0)+2*(0,-\rowsp)$) (ca_label){$\rm \bf Ca^+$};
\draw[qubit_line] ($-4.*(\colsp,0)$) -- +($0.5*(\colsp,0)$);
\draw[qubit_line] ($-4*(\colsp,0)+(0,-\rowsp)$) -- +($0.5*(\colsp,0)$);
\draw[qubit_line] ($-4*(\colsp,0)+2*(0,-\rowsp)$) -- +($0.5*(\colsp,0)$);

\node at  ($-3.1*(\colsp,0)-1*(0,\rowsp)$) (leftellipsis) {\bf \large \ldots};
\draw[qubit_line] ($-2.8*(\colsp,0)$) -- +($8.5*(\colsp,0)$);
\draw[qubit_line] ($-2.8*(\colsp,0)+(0,-\rowsp)$) -- +($8.5*(\colsp,0)$);
\draw[qubit_line] ($-2.8*(\colsp,0)+2*(0,-\rowsp)$) -- +($8.5*(\colsp,0)$);

\node[qnd_block](Um) at ($-2*(\colsp,0)-1*(0,\rowsp)$)  {$\bf U_{S_Z}$};
\pic at ($-1.1*(\colsp,0)-2*(0,\rowsp)$) {measCa};
\node[sqCa] at ($-0.1*(\colsp,0)-2*(0,\rowsp)$) {cool,\\init};
\node[qnd_block](Um) at ($1*(\colsp,0)-1*(0,\rowsp)$)  {$\bf U_{S_X}$};
\pic at ($2*(\colsp,0)-2*(0,\rowsp)$) {measCa};
\node[sqCa] at ($2.9*(\colsp,0)-2*(0,\rowsp)$) {cool,\\init};
\node[feedback_block](feed_z) at ($3.7*(\colsp,0)-0.5*(0,\rowsp)$)  {$\rm \bf C_z$};
\node[feedback_block](feed_x) at ($4.7*(\colsp,0)-0.5*(0,\rowsp)$)  {$\rm \bf C_x$};

\node[inner sep=0,minimum size=0] at ($-1.1*(\colsp,0)-2.5*(0,\rowsp)$) (meas_z) {}; 
\node[inner sep=0,minimum size=0] at ($-1.1*(\colsp,0)-3.3*(0,\rowsp)$) (int_z) {}; 
\node[inner sep=0,minimum size=0] at ($-1.1*(\colsp,0)-3.3*(0,\rowsp)$) (int2_z) {}; 
\draw[vecArrow] (meas_z) |- (int_z)to (int2_z) -|(feed_z);
\node[inner sep=0,minimum size=0] at ($2*(\colsp,0)-2.5*(0,\rowsp)$) (meas_x) {}; 
\node[inner sep=0,minimum size=0] at ($2*(\colsp,0)-3.0*(0,\rowsp)$) (int_x) {}; 
\node[inner sep=0,minimum size=0] at ($2*(\colsp,0)-3.0*(0,\rowsp)$) (int2_x) {}; 
\draw[vecArrow] (meas_x) |- (int_x)to (int2_x) -|(feed_x);
\draw (-2.3,+0.5) -- (4.6,0.5);
\draw (-2.3,+0.5) -- (-2.3,0.4);
\draw (4.6,+0.5) -- (4.6,0.4);
\node at ($1.0*(\colsp,0)+1*(0,\rowsp)$) {$\bf i$};
  
%
%
%
%
\node at  ($6.3*(\colsp,0)-1*(0,\rowsp)$) (leftellipsis) {\bf \large \ldots};

\end{tikzpicture}}
  \includegraphics[width = \columnwidth]{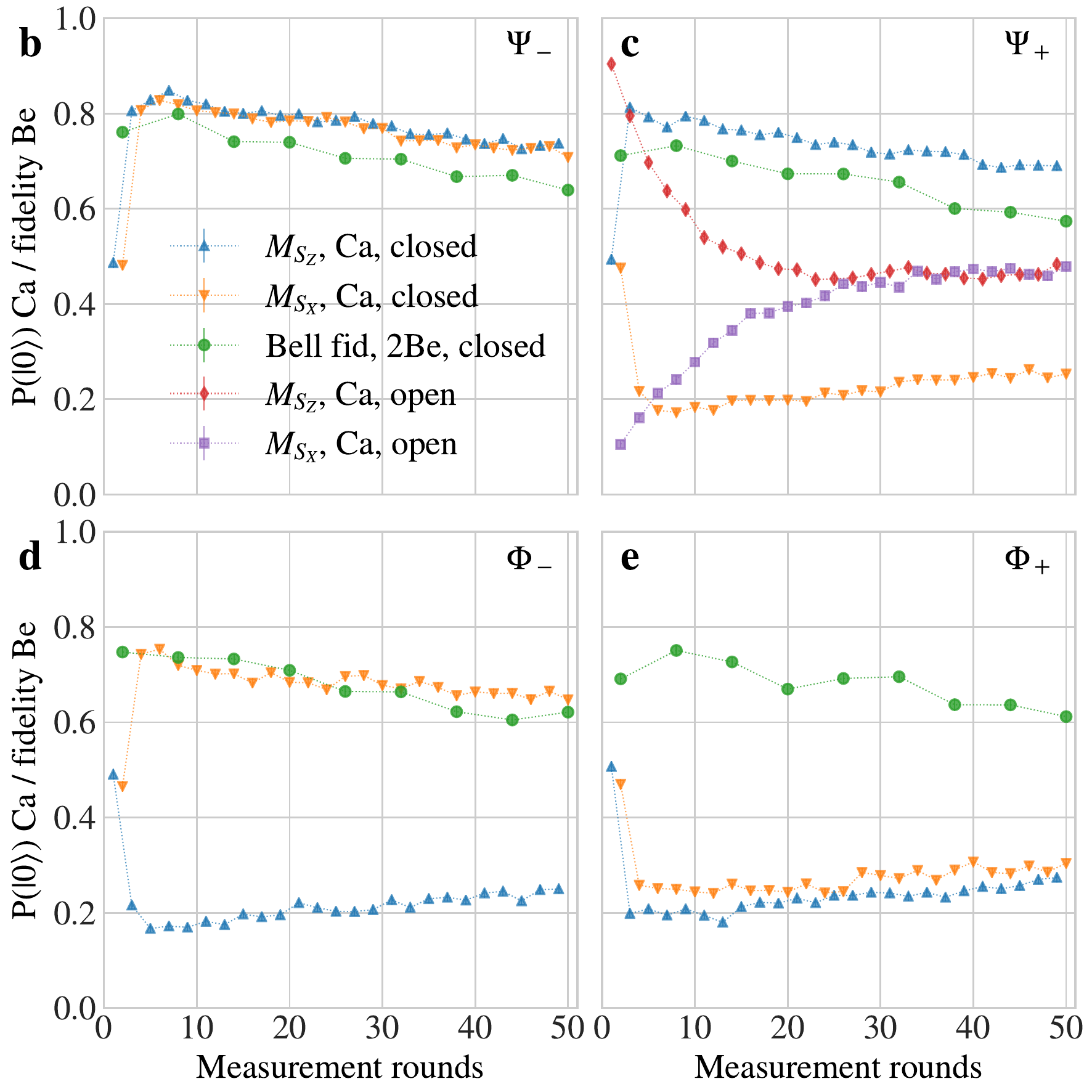}
  \caption{Bell state stabilization.
    \textbf{a.} Sequence for a measurement and feedback cycle used to generate and stabilize Bell states.
    $M_{S_Z}$ and $M_{S_X}$ are performed sequentially, followed by sequential conditional feedback $C_Z$ and $C_X$.
    \textbf{b -- e.} Evolution of the measured \Caplus outcome probabilities over 25 measurement and feedback cycles, which each contain 50 rounds of parity measurement.
    Also shown are the fidelities for the relevant states measured after stopping the sequence at a fixed point and performing a set of measurements directly on beryllium.
    In \textbf{c} we also show the results from an open loop experiment performed using the input state $\ket{\Phi_+}$, and repeatedly applying $M_{S_Z}$ without any feedback.
    The red diamonds show measurements of $S_{Z}$, while the purple squares are for $S_X$.
    Each data point is the result of 2000 measurements, resulting in statistical uncertainties which are smaller than markers.
  }
  \label{fig:bell_state_prep}
\end{figure}
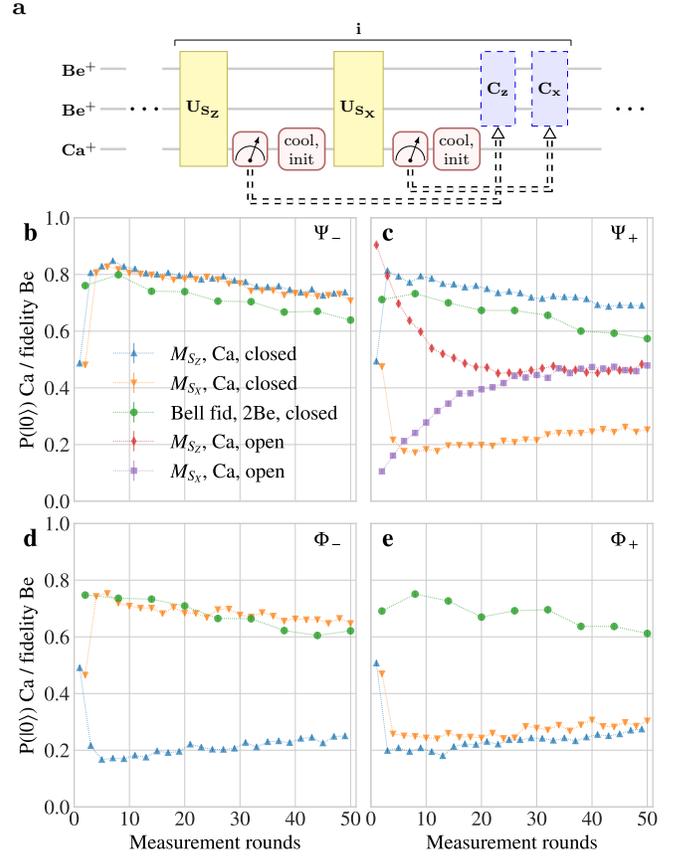
We now proceed to stabilize the four Bell states $(\Phi_\pm = (\ket{00}\pm \ket{11})/\sqrt{2}$, $\Psi_\pm = (\ket{01}\pm \ket{10})/\sqrt{2})$ of two qubits, which are stabilized by the two commuting operators $S_Z$ and $S_X$.
These states can thus be prepared starting from an arbitrary input by sequentially measuring $S_Z$ and $S_X$ and performing conditional feedback if the measured eigenvalues differ from the desired result.
We apply the correction operations $C_Z$, $C_X$ after both stabilizers have been read out.
Results for stabilization of these states are shown in Fig.\,\ref{fig:bell_state_prep}.
Also shown are the Bell state fidelities measured by sampling the state at fixed points in the sequence.
These are determined using a combination of coherent rotations and subsequent measurements performed directly on the \Beplus ions (see Methods).
As a comparison with the stabilized states, we also show data for an unstabilized $\Psi_+$ Bell state input, which was produced by a single two-qubit MS gate on the \Beplus ions followed by a $R_{\pi/2}(\phi_p)$ pulse on both qubits with the appropriate phase.
We observe a mean fidelity after a single block of measuring both stabilizers and correction for all of the Bell states of 0.731(4), which drops to a mean of $0.613(4)$ after 25 cycles (since each cycle contains two stabilizer measurements, this is 50 measurement rounds), thus still exhibiting that entanglement is preserved in the system even after this extended sequence of operations.
The highest fidelity achieved is for the singlet $\ket{\Psi_-}$ state, while the lowest is for $\ket{\Phi_+}$.
This may be due to noise or an offset in the frequency calibration of the qubit -- the singlet state is insensitive to common mode phase errors, while the $\ket{\Phi_+}$ state is most sensitive to this source.

Our experiments show the general elements of stabilizer readout and correction, as required for performing quantum error correction.
Nevertheless many improvements are needed in order for QEC to operate reliably.
QEC requires measurement of higher weight stabilizer operators \cite{bombin_topological_2006, fowler_surface_2012, nigg_quantum_2014}, and needs to be made compatible with the demands of fault-tolerance \cite{gottesman_theory_1998}.
In our system the primary source of error is the stabilizer readout operation itself; the qubit coherence time of our beryllium qubits is longer than any sequence we demonstrate.
Of primary importance is therefore to reduce errors in the basic operations.
In our system, these stem from motional decoherence during the multi-species gates, decoherence of the calcium qubit during the measurement block, and pulse calibration errors.
We find that combining many different elements including feedback operations applied probabilistically during the sequence adds significant complexity to the control and debugging of these systems, which requires an increased level of automation.

The control and measurement demonstrated here offer direct access to temporal correlations of multi-qubit systems (see Methods), with the potential to provide new insights into the evolution of quantum systems.
The addition of conditional feedback opens up number of opportunities for quantum state control.
While in this work we have focussed on QEC, alternative applications include investigations of measurement-based quantum computing \cite{raussendorf_one-way_2001, lanyon_measurement-based_2013}, quantum gate teleportation \cite{gottesman_demonstrating_1999} or quantum metrology \cite{leibfried_toward_2004}.

We thank Daniel Kienzler, Ludwig de Clercq and David Nadlinger for important contributions to the apparatus, and Joseba Alonso for useful discussions.
We acknowledge support from the Swiss National Science Foundation through the National Centre of Competence in Research for Quantum Science and Technology (QSIT) grant 51NF40-160591.
KM is supported by an ETH Z\"urich Postdoctoral Fellowship.
The research is partly based upon work supported by the Office of the Director of National Intelligence (ODNI), Intelligence Advanced Research Projects Activity (IARPA), via the U.S. Army Research Office grant W911NF-16-1-0070.

The views and conclusions contained herein are those of the authors and should not be interpreted as necessarily representing the official policies or endorsements, either expressed or implied, of the ODNI, IARPA, or the U.S. Government.
The U.S. Government is authorized to reproduce and distribute reprints for Governmental purposes notwithstanding any copyright annotation thereon.
Any opinions, findings, and conclusions or recommendations expressed in this material are those of the author(s) and do not necessarily reflect the view of the U.S. Army Research Office. 

\appendix*
\section{\Large Methods}

\textbf{B-field insensitive beryllium qubit.}
Beryllium qubits are stored in the $\ket{0} \equiv \ket{F=1, m_F=+1}$ and $\ket{1} \equiv \ket{F=2, m_F=0}$ states of the ground-state hyperfine structure.
The transition frequency is insensitive to first-order in the magnetic field strength for fluctuations around the 119.2 G applied B-field used in our experiment.
As a result, we obtain coherence times of over 1 second for a single qubit.
Internal state preparation of the beryllium ions is performed by optical pumping to the $\ket{F=2, m_F=2}$ state, followed by a coherent Raman transition which transfers population to $\ket{0}$.
Following the experimental sequence described in the main text, direct readout of beryllium is performed by first using two Raman transfer pulses  $\ket{0} \rightarrow \ket{F=2, m_F=2}$ and $\ket{1} \rightarrow \ket{F=1, m_F=-1}$, and subsequently illuminating the ions with resonant light on the $\ket{S_{1/2},F=2, m_F=2} \rightarrow \ket{P_{3/2},F=3, m_F=3}$ transition \cite{de_clercq_parallel_2016}.
Photons scattered by the ion are detected on a photo-multiplier tube, resulting in distributions with a mean of 25 (0.2) counts for the bright (dark) state.
Due to impure polarization causing optical pumping, significant departure from Poissonian distributions is seen.
This results in an error of around 0.5\% per ion in state detection.

\begin{figure}[t]
  \includegraphics[width = .8\columnwidth]{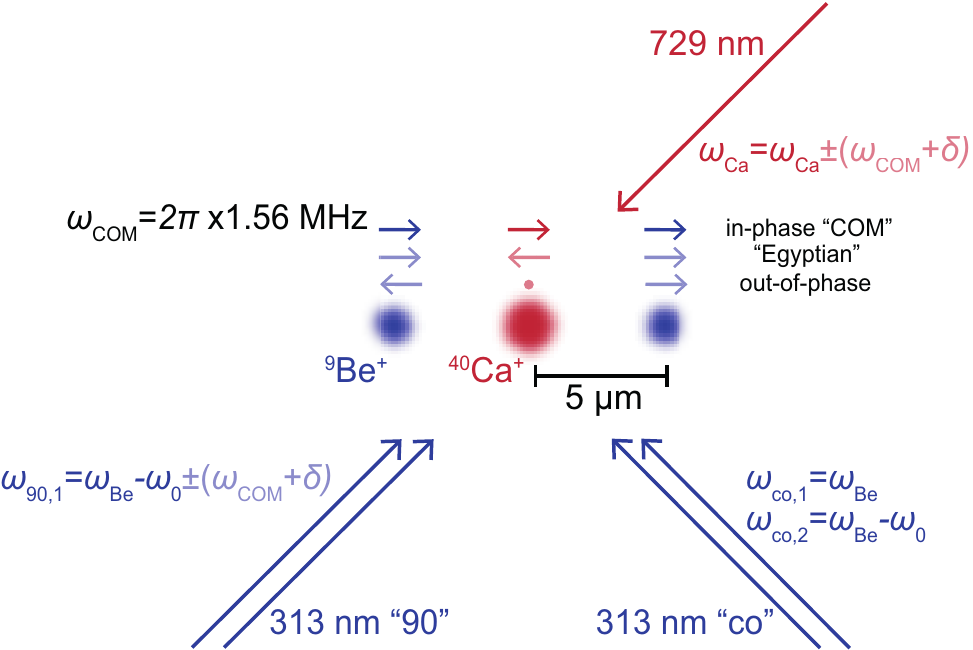}
  \caption{Ion crystal and beam configuration.
    3 co-trapped ions in a single harmonic potential well are addressed by Raman beam pairs for \Beplus and a 729 beam for \Caplus.
    The laser frequencies written in solid text correspond to those applied for single-qubit operations, resonant with the qubit transition; and the faint text represents modified frequencies used to address the sidebands associated with the in-phase center-of-mass (COM) mode of the 3-ion crystal at 1.56\,MHz, at a small detuning $\delta$ to drive the entangling MS gates. Arrows above the ions indicate the normal-mode displacement directions for this mode, as well as the ``Egyptian" (4.20 MHz) and the out-of-phase (4.11 MHz) axial modes.}
  \label{fig:beams}
\end{figure}

\textbf{ Ion crystal and normal modes.}
Experiments are performed in a segmented linear Paul trap with an ion-electrode distance of 185\,$\mu$m.
The ion chain used in the experiments is a 3-ion chain (\Beplus-\Caplus-\Beplus) with calcium in the centre, which is formed and maintained using standard re-ordering techniques after every 50 experimental sequences \cite{home_chapter_2013}.
This allows the ions to recover from occasional re-ordering events occurring due to background gas collisions.
The ions exhibit 3 axial normal modes, schematically illustrated in Fig.\,\ref{fig:beams}.
For a harmonic potential, the calcium ion is decoupled from the mode in which the beryllium ions oscillate out of phase, which makes it unsuitable for multi-species gate operations.
We choose the lowest-frequency motional mode (with a frequency of 1.56\,MHz) in which all ions oscillate in-phase for the MS gates used in the parity measurement.
For this mode, the Lamb-Dicke parameter is 0.13 for the beryllium ions and 0.07 for the calcium ion.
We cool the crystal before each experimental run using first 3\,ms of far-detuned and then 600\,$\mu$s of near-resonant Doppler cooling on \Beplus and \Caplus simultaneously.
This is followed by \Caplus EIT cooling optimised on the lowest radial mode at 2.5\,MHz, \Beplus Raman sideband cooling on the two highest-frequency axial motional modes, and finally \Caplus sideband cooling on the lowest-frequency mode.

Once every several minutes the ions tend to de-crystallize, resulting in a loss of fluorescence and the possible loss of an ion without rapid intervention.
This is mitigated by monitoring the fluorescence emitted by the \Beplus ions during Doppler cooling for each experimental shot.
If it is below a threshold for more than a few shots, then the trap axial frequency is immediately reduced to ~400\,kHz, resulting in stronger radial confinement, and a loop of far-detuned and near-resonant Doppler cooling is repeated until the fluorescence returns to normal or a timeout is reached.
In the former case the experiment is seamlessly re-run without user intervention.
The loop takes 50--200\,ms to complete.
With automatic re-crystallization a 3-ion crystal can remain trapped for 4--6 hours at a time, with beryllium hydride formation being the dominant cause of ion loss.

\textbf{Qubit control: beryllium and calcium.}
\Beplus qubit operations are mediated by pairs of Raman beams near-resonant with the 1.2 GHz qubit transition frequency.
For single-qubit operations, we use a pair of beams which are co-propagating as they enter the trap (the ``co" beam pair in Fig.\,\ref{fig:beams}), such that any path-length fluctuations are common to both and thus the difference phase, which sets the phase in the qubit frame of reference, exhibits minimal drift.
The wavevector difference between these two beams is nearly zero, which means that they cannot mediate spin-motion coupling.
For this reason, a second beam path involving a Raman beam pair entering the apparatus at 90 degrees to one another with their difference wavevector aligned along the trap axis is used for the MS entangling gates.
Owing to the different optical paths, air currents and slight thermal drifts result in submicron-scale path-length fluctuations which are not common to both beams, and hence the phase $\phi_b$ discussed in the main text can drift.
This is a known problem with such free-space beam geometries \cite{lee_phase_2005}.
We take advantage of the fact that these drifts are slow on the 100 $\mu$s timescales associated with our individual unitaries to enclose the gates using the ``90'' beam pair (the MS gates and \Beplus $\pi$ pulse) in a Ramsey sequence using the same pair, which serves to cancel this phase sensitivity.

The calcium qubit states are $\ket{0} \equiv \ket{^2 S_{1/2}, m_J=+\frac{1}{2}}$ and $\ket{1} \equiv \ket{^2 D_{5/2}, m_J=+\frac{3}{2}}$, for which the transition frequency is first-order dependent on the magnetic field, resulting in a much shorter coherence time of 1.5 ms.
Internal state preparation of calcium ions is performed by frequency selective optical pumping \cite{kienzler_quantum_2015}.
The \Caplus qubit is controlled using a single 729\,nm laser propagating 45$^\circ$ with respect to the trap axis.


\textbf{Phase updates and reset.}
Laser pulses, including frequency, amplitude and phase are controlled by radio-frequency synthesis \cite{keitch_programmable_2017}.
Each individual frequency used in the experiment is initialized to zero phase at a certain point prior to the coherent operations performed on the ions.
The phase of each pulse is then referenced to a `reference' time $t_{start}$.
Phases for every gate are calculated according to $\phi=\omega_{rf} (t-t_{start})$, such that carrier pulses remain in phase with the qubit.
While this scheme ensures that all single-frequency pulses from a particular beam or beam pair remain phase-coherent throughout an experiment, it cannot preserve the relative phase relationship between rf pulses at different frequencies, such as between carrier rotations and MS gates, when the whole block of pulses is shifted in time.
Such a shift occurs for the stabilizer readout block of pulses since it repeats throughout the sequence.
To mitigate this the reference time $t_{start}$ is shifted to a fixed time prior to every $U_{S_Z}$ stabilizer readout pulse block on each occasion that the latter is applied.
It is then reset back to the experiment start time prior to any further carrier rotations on the qubits.
Since the $S_Z$ stabilizer readout block is diagonal in the computational basis, the phases of the pulses in this block do not need to be referenced to the rest of the sequence.

In addition to the time reference shifting, the Stark shifts due to $U_{S_Z}$, $U_{S_X}$, $C_X$ and $C_Z$ operations in the feedback experiments are dynamically compensated by adjusting future gate phases based on previously-conducted operations.
Unlike the phase accumulation described earlier, which is hard-coded, these calculations are carried out rapidly in our embedded software and the results are pushed to the rf hardware with low latency.

\textbf{Fidelity estimation of single parity measurements.}
For a single round of parity state preparation and measurement, the measurement fidelity $F_{\rm ND}$ quoted in the main text is an estimate of how well the \Caplus readout probabilities agree with the \Beplus input state parity probability distribution.
The quantum state preparation fidelity $F_{\rm QSP}$ gives a measure of the mean conditional probability that the resulting \Beplus subspace corresponds to the measurement outcome obtained from the \Caplus readout \cite{ralph_quantum_2006, barreiro_open-system_2011}.
Mathematically these are given by 
\begin{align*}
  F_{\rm ND} &= \left( \sqrt{p^{\rm in}_{+1}\,p^{\rm m}_{\ket{1}}} + \sqrt{p^{\rm in}_{-1}\,p^{\rm m}_{\ket{0}}} \right)^2 \\
  F_{\rm QSP} &= p^{\rm m}_{\ket{1}}\,p^{\rm out}_{\ket{1},+1} + p^{\rm m}_{\ket{0}}\,p^{\rm out}_{\ket{0},-1}
\end{align*}
where $p^{\rm in}_{+1/-1}$ is the probability to find the two \Beplus ions in the  $E_Z = +1$ ($E_Z = -1$) subspace based on the direct measurement performed on the input state (violet stars in Fig.\,1b of the main text show $p^{\rm in}_{+1}$).
$p^{\rm m}_{\ket{1}}$ ($p^{\rm m}_{\ket{0}}$) is the probability of measuring \Caplus in the dark (bright) state. $p^{\rm out}_{\ket{1},+1}$ ($p^{\rm out}_{\ket{0},-1}$) is the conditional probability that the measurement of the parity directly performed on the beryllium ions after $M_Z$ corresponds to the measurement result.
$F_{\rm QSP}$ is the most relevant for feedback experiments.
$\overline{F}_{\rm ND}$ and $\overline{F}_{\rm QSP}$ are averages of $F_{\rm ND}$ and $F_{\rm QSP}$ over all input state preparation phases.

\textbf{Direct estimation of beryllium Bell state fidelity.} To estimate beryllium state fidelities we measure correlations between the ions in three orthogonal bases.
The first is the computational basis, which we obtain by a direct measurement of the two ions, extracting the probability to find the ions in the same state which we use to obtain $\expect{ZZ}$.
For the $\Phi_\pm$ states, we then apply a $R_{\pi/2}$ pulse to both ions with the phase of the pulse on both ions set to measure either $XX$ or $YY$.
Correlations observed in the subsequent fluorescence detections then allow us to estimate $\expect{XX}$ and $\expect{YY}$.
For the $\Psi_\pm$ states, we precede the common $R_{\pi/2}$ pulse with a $C_Z = -I_1\otimes X_2$ operation, which converts the states to one of the $\Phi_\pm$ states, with the resulting analysis performed in the same manner as before.
While the use of the $C_Z$ operation is not strictly necessary, it provides a simple set of diagnostics which we find easy to manage and debug.
In the feedback experiments, it is important to ensure that the phase of the pulses used in the analysis is updated dependent on the beryllium feedback operations which have been performed. 

\textbf{Correlations in the Bell state stabilization.}
To better understand the Bell state stabilization data, we investigated the two-point correlations between successive \Caplus readouts in the $S_Z$ or $S_X$ bases for the four Bell states, which provided useful diagnostic information on the feedback operations.

\begin{figure}[tb] 
\includegraphics[width=\columnwidth]{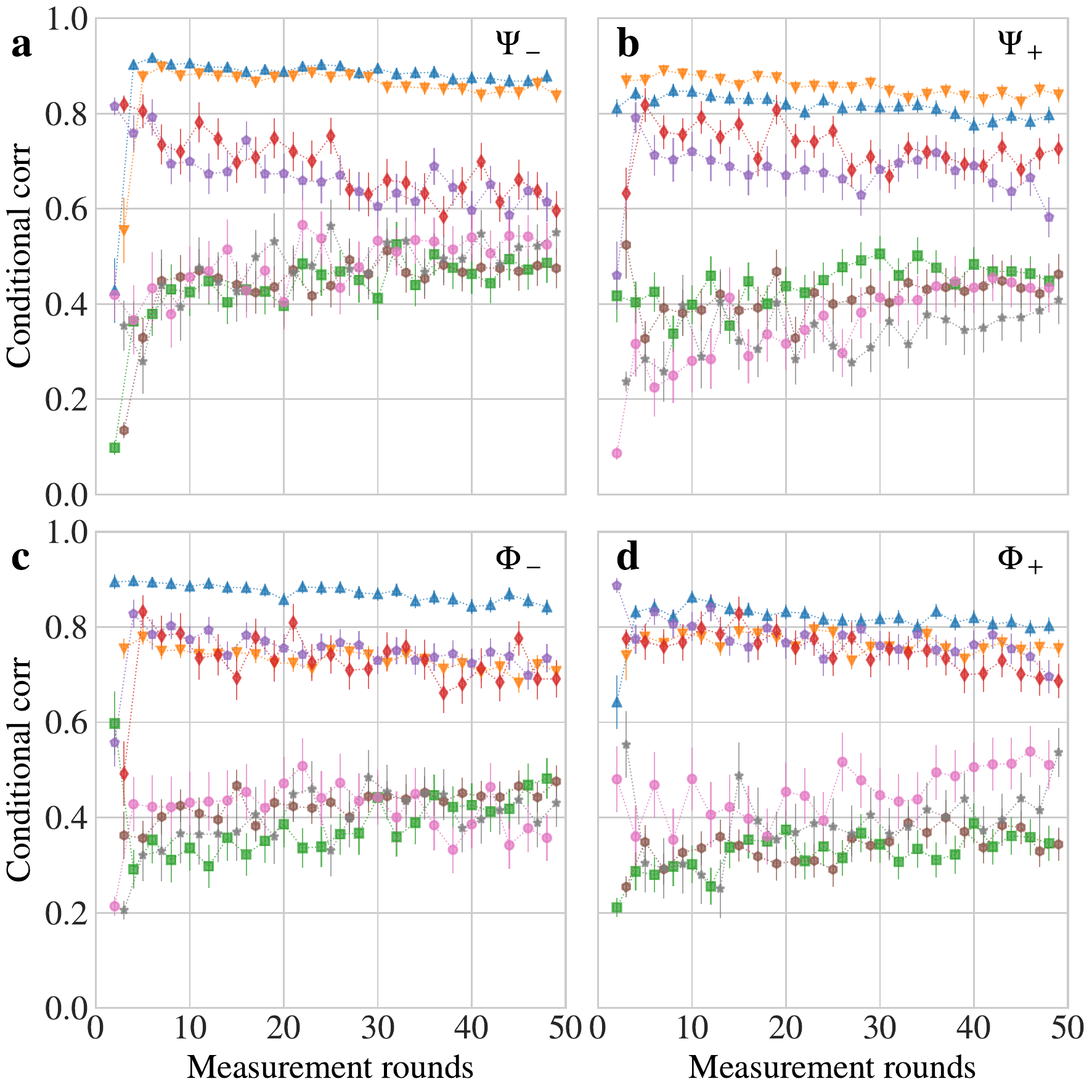}

  \caption{Experimental Bell measurement correlations.
    Correlations between successive $M_{S_Z}$ and $M_{S_Z}$ measurement operations categorised by the feedback that took place between them.
    Values of 1 (0) represent perfect (anti-) correlation.
    Upward- (downward)- pointing triangles show the $S_Z$ ($S_X$) correlations where no feedback occurred, thus for $S_Z$ the sequence is $S_Z - S_X - S_Z$.
    Pentagons (diamonds) are data where $C_X$ ($C_Z$) was applied, which nominally commutes with $S_Z$ ($S_X$).
    One example would be a sequence $S_Z - S_X - C_X - S_Z$.
    Squares (hexagons) are correlations between two measurements where $C_Z$ ($C_X$) took place, namely a correction which anti-commutes with the operator of interest.
    Circles (stars) indicate data where both $C_X$ and $C_Z$ occurred.
    The plots use the same raw data as Fig.\,\ref{fig:bell_state_prep}, and uncertainties reflect projection noise; they fluctuate significantly due to the varying numbers of events for each point, which occurs since each point corresponds to a subset of the full data.
  }
  \label{fig:bell_correlations}
\end{figure}

Each correlation datum was categorised based on which correction operations had taken place between the two stabilizer measurements, and the subdivided data is shown in Fig.\,\ref{fig:bell_correlations}.
Ideally, for $S_Z$ ($S_X$), these correlations would have value 1 for a $C_X$ ($C_Z$) operation, since it commutes with the stabilizer.

We observe that the correlations are on average $\approx 10$\% lower with nominally-commuting correction operations than the case without feedback, which indicates that experimentally the $C_X$ ($C_Z$) operations did not commute with $U_{S_Z}$ ($U_{S_X}$) as well as expected from calibration experiments, where errors on the level of 1\% would be expected.

In a perfect implementation of measurement and feedback, then aside from the first round of correction, the $C_Z$ correction operation is only applied when an additional source of error changes the subspace.
In this case, the $S_Z$ correlation between two detections with a $C_Z$ operation between them should always be anti-correlated.
However in the case where the dominant error is due to imperfect readout, the steady state situation produces a 50\% anti-correlation, since in half of all cases the correction operation is applied on a state which is uncorrupted.
Our data is primarily affected by errors during the readout, thus for this correlation we see higher levels of between 30--50\%.

\textbf{Simulation and modelling.}
We simulate the behaviour of our sequence of measurements using a simple model which allows us to quantify the main observed behaviours.
This is implemented in a Monte-Carlo simulation of the density matrix.
The action of the measurement sequence itself is modelled by the ideal unitary operators $U_{S_{Z/X}}$ that map the parity information of the data qubits onto the ancilla.
We mimic decoherence and Raman off-resonant scattering using two additional maps.
The first is a depolarizing channel \cite{nielsen_quantum_2000}, which is applied as a map to the density operator after the application of the unitary, but before the measurement result is extracted.
This is parameterized by a depolarization rate per measurement round of $\gamma_{dep}$ .
The measurement of the calcium ion is then performed using a projection.
Following this, we add a leakage channel, which is implemented by a partial trace of one or both of the beryllium ions applied probabilistically based on comparison of a random number with a reference drawn from the leakage probability distribution.
This uses a leakage decay rate per ion of $\gamma_{leak}$ per measurement round.
We fix the leakage rate to $\gamma_{leak} = 0.3\%$ which was experimentally determined with independent measurements of the population remaining in the qubit subspace after applying the looped sequence with just one of the two Raman beams on at a time.
The depolarization rate $\gamma_{dep}$ is instead adjusted for each data set.
In the open-loop data we adjust $\gamma_{dep}$ to best reproduce the exponential decay of the state while  for the closed-loop sets we use it to adjust the offset of the first few points thus reflecting the infidelities of the feedback operations.
We find for the open-loop $S_Z$ measurement $\gamma_{dep}=0.06$ ($\gamma_{dep}=0.07$) for $E_Z=-1$ ($E_Z=+1$) and $\gamma_{dep}=0.11$ for $E_X=+1$.
We interpret the higher value for $E_X=+1$ measurements as due to the higher susceptibility to phase fluctuations.
For the closed-loop sets instead we find $\gamma_{dep}=0.10$ ($\gamma_{dep}=0.10$) for $E_Z=-1$ ($E_Z=+1$) and $\gamma_{dep}=0.15$ ($\gamma_{dep}=0.16$) for $E_X=+1$ ($E_X=-1$). These higher values are connected to imperfections in the correction pulses as described in the previous section.  We note that the errors occuring in our experimental system are most likely not well modelled by depolarizing noise. Nevertheless in this complex system it gives a simple single-parameter adjustment which accounts for decoherence.

As can be seen in Fig.\,\ref{fig:subspace_stab} of the main text, the simulation results including only leakage do not fully reproduce the observed dynamics.
By adding a gradual drift in the readout fidelity of the parity measurement, we are able to simultaneously produce a good match to the data for all datasets.
Such a drift could arise experimentally due to a number of sources.
We observe gradual changes in pulse amplitudes over the sequence length, as well as a change in the photo-multiplier-tube counts.
Combined, these problems are consistent with a parity detection bias that increases over the duration of the sequence to $3-5\%$.

\begin{figure}[tb]
	\includegraphics[width = \columnwidth]{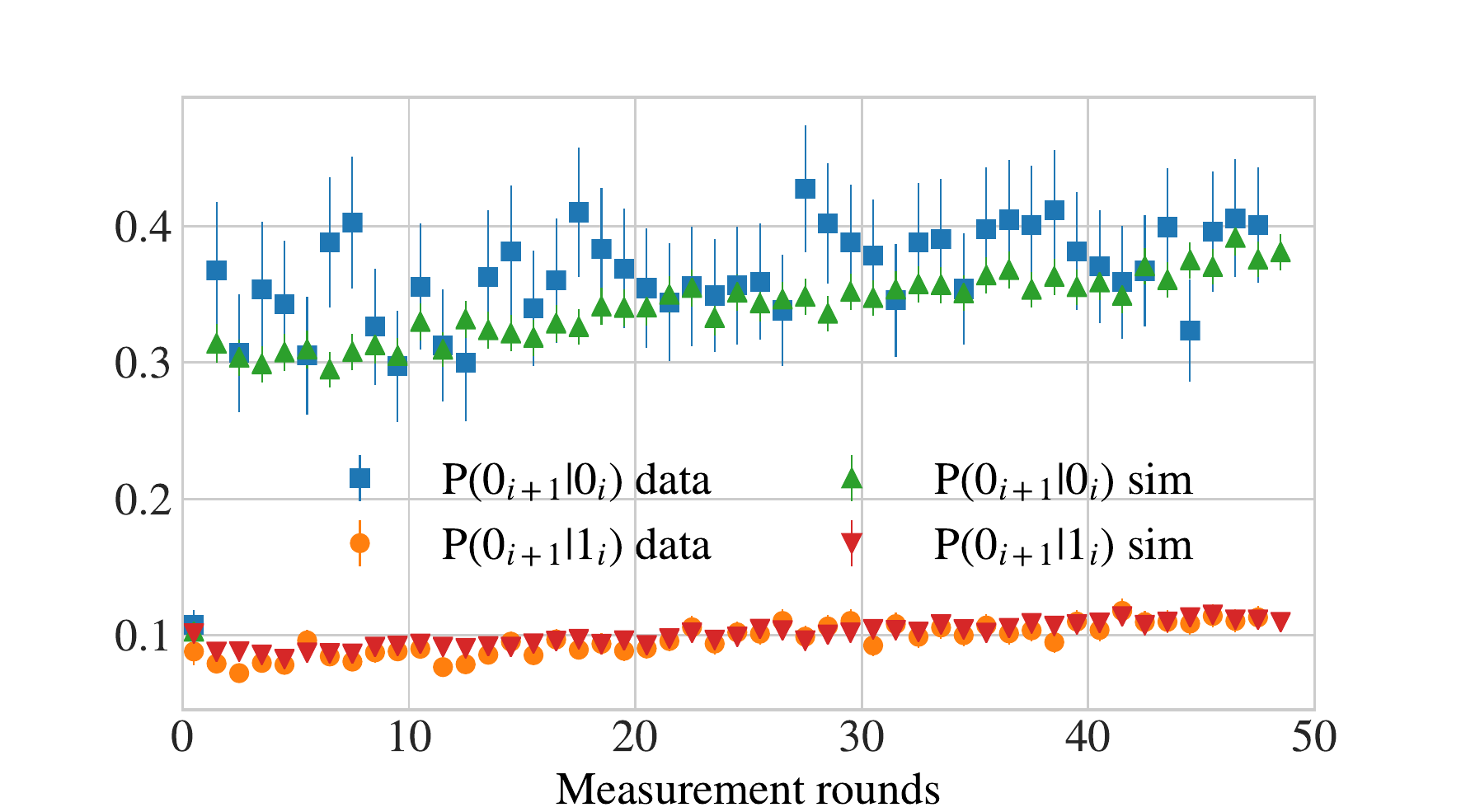}
	\caption{Simulated subspace measurement correlations.
          Defining $P(0_i)$ as the probability of applying a feedback operation in the $i$-th measurement round, we plot the conditional probability of feeding back on the system twice in a row ($P(0_{i+1}|0_{i})$) or just once ($P(0_{i+1}|1_{i})$) for both the experimental data and the simulation.
          Uncertainties for $P(0_{i+1}|0_{i})$ are much larger due to the rarity of these events (an average of 100 events over 10\,000 simulated points).}
	\label{fig:correlation_sim}
\end{figure}

Using this simulation  we can also examine the behaviour of correlations between feedback events as described in the previous section.
In Fig.\,\ref{fig:correlation_sim} we plot the probabilities that two feedback events are applied in subsequent shots from both simulation and theory. We notice that the simulation reproduces the results observed in the data reasonably well, which verifies the arguments made regarding correlations in the previous section.

\bibliography{ETHZ}

\begin{thebibliography}{42}%
\makeatletter
\providecommand \@ifxundefined [1]{%
 \@ifx{#1\undefined}
}%
\providecommand \@ifnum [1]{%
 \ifnum #1\expandafter \@firstoftwo
 \else \expandafter \@secondoftwo
 \fi
}%
\providecommand \@ifx [1]{%
 \ifx #1\expandafter \@firstoftwo
 \else \expandafter \@secondoftwo
 \fi
}%
\providecommand \natexlab [1]{#1}%
\providecommand \enquote  [1]{``#1''}%
\providecommand \bibnamefont  [1]{#1}%
\providecommand \bibfnamefont [1]{#1}%
\providecommand \citenamefont [1]{#1}%
\providecommand \href@noop [0]{\@secondoftwo}%
\providecommand \href [0]{\begingroup \@sanitize@url \@href}%
\providecommand \@href[1]{\@@startlink{#1}\@@href}%
\providecommand \@@href[1]{\endgroup#1\@@endlink}%
\providecommand \@sanitize@url [0]{\catcode `\\12\catcode `\$12\catcode
  `\&12\catcode `\#12\catcode `\^12\catcode `\_12\catcode `\%12\relax}%
\providecommand \@@startlink[1]{}%
\providecommand \@@endlink[0]{}%
\providecommand \url  [0]{\begingroup\@sanitize@url \@url }%
\providecommand \@url [1]{\endgroup\@href {#1}{\urlprefix }}%
\providecommand \urlprefix  [0]{URL }%
\providecommand \Eprint [0]{\href }%
\providecommand \doibase [0]{http://dx.doi.org/}%
\providecommand \selectlanguage [0]{\@gobble}%
\providecommand \bibinfo  [0]{\@secondoftwo}%
\providecommand \bibfield  [0]{\@secondoftwo}%
\providecommand \translation [1]{[#1]}%
\providecommand \BibitemOpen [0]{}%
\providecommand \bibitemStop [0]{}%
\providecommand \bibitemNoStop [0]{.\EOS\space}%
\providecommand \EOS [0]{\spacefactor3000\relax}%
\providecommand \BibitemShut  [1]{\csname bibitem#1\endcsname}%
\let\auto@bib@innerbib\@empty
\bibitem [{\citenamefont {Steane}(1996)}]{steane_error_1996}%
  \BibitemOpen
  \bibfield  {author} {\bibinfo {author} {\bibfnamefont {A.~M.}\ \bibnamefont
  {Steane}},\ }\href {\doibase 10.1103/PhysRevLett.77.793} {\bibfield
  {journal} {\bibinfo  {journal} {Physical Review Letters}\ }\textbf {\bibinfo
  {volume} {77}},\ \bibinfo {pages} {793} (\bibinfo {year} {1996})}\BibitemShut
  {NoStop}%
\bibitem [{\citenamefont {Shor}(1995)}]{shor_scheme_1995}%
  \BibitemOpen
  \bibfield  {author} {\bibinfo {author} {\bibfnamefont {P.~W.}\ \bibnamefont
  {Shor}},\ }\href {\doibase 10.1103/PhysRevA.52.R2493} {\bibfield  {journal}
  {\bibinfo  {journal} {Physical Review A}\ }\textbf {\bibinfo {volume} {52}},\
  \bibinfo {pages} {R2493} (\bibinfo {year} {1995})}\BibitemShut {NoStop}%
\bibitem [{\citenamefont {Gottesman}(1998)}]{gottesman_theory_1998}%
  \BibitemOpen
  \bibfield  {author} {\bibinfo {author} {\bibfnamefont {D.}~\bibnamefont
  {Gottesman}},\ }\href {\doibase 10.1103/PhysRevA.57.127} {\bibfield
  {journal} {\bibinfo  {journal} {Physical Review A}\ }\textbf {\bibinfo
  {volume} {57}},\ \bibinfo {pages} {127} (\bibinfo {year} {1998})}\BibitemShut
  {NoStop}%
\bibitem [{\citenamefont {Leibfried}\ \emph {et~al.}(2004)\citenamefont
  {Leibfried}, \citenamefont {Barrett}, \citenamefont {Schaetz}, \citenamefont
  {Britton}, \citenamefont {Chiaverini}, \citenamefont {Itano}, \citenamefont
  {Jost}, \citenamefont {Langer},\ and\ \citenamefont
  {Wineland}}]{leibfried_toward_2004}%
  \BibitemOpen
  \bibfield  {author} {\bibinfo {author} {\bibfnamefont {D.}~\bibnamefont
  {Leibfried}}, \bibinfo {author} {\bibfnamefont {M.~D.}\ \bibnamefont
  {Barrett}}, \bibinfo {author} {\bibfnamefont {T.}~\bibnamefont {Schaetz}},
  \bibinfo {author} {\bibfnamefont {J.}~\bibnamefont {Britton}}, \bibinfo
  {author} {\bibfnamefont {J.}~\bibnamefont {Chiaverini}}, \bibinfo {author}
  {\bibfnamefont {W.~M.}\ \bibnamefont {Itano}}, \bibinfo {author}
  {\bibfnamefont {J.~D.}\ \bibnamefont {Jost}}, \bibinfo {author}
  {\bibfnamefont {C.}~\bibnamefont {Langer}}, \ and\ \bibinfo {author}
  {\bibfnamefont {D.~J.}\ \bibnamefont {Wineland}},\ }\href {\doibase
  10.1126/science.1097576} {\bibfield  {journal} {\bibinfo  {journal}
  {Science}\ }\textbf {\bibinfo {volume} {304}},\ \bibinfo {pages} {1476}
  (\bibinfo {year} {2004})}\BibitemShut {NoStop}%
\bibitem [{\citenamefont {Terhal}(2015)}]{terhal_quantum_2015}%
  \BibitemOpen
  \bibfield  {author} {\bibinfo {author} {\bibfnamefont {B.~M.}\ \bibnamefont
  {Terhal}},\ }\href {\doibase 10.1103/RevModPhys.87.307} {\bibfield  {journal}
  {\bibinfo  {journal} {Reviews of Modern Physics}\ }\textbf {\bibinfo {volume}
  {87}},\ \bibinfo {pages} {307} (\bibinfo {year} {2015})}\BibitemShut
  {NoStop}%
\bibitem [{\citenamefont {Steane}(2003)}]{steane_overhead_2003}%
  \BibitemOpen
  \bibfield  {author} {\bibinfo {author} {\bibfnamefont {A.~M.}\ \bibnamefont
  {Steane}},\ }\href {\doibase 10.1103/PhysRevA.68.042322} {\bibfield
  {journal} {\bibinfo  {journal} {Physical Review A}\ }\textbf {\bibinfo
  {volume} {68}},\ \bibinfo {pages} {042322} (\bibinfo {year}
  {2003})}\BibitemShut {NoStop}%
\bibitem [{\citenamefont {Cory}\ \emph {et~al.}(1998)\citenamefont {Cory},
  \citenamefont {Price}, \citenamefont {Maas}, \citenamefont {Knill},
  \citenamefont {Laflamme}, \citenamefont {Zurek}, \citenamefont {Havel},\ and\
  \citenamefont {Somaroo}}]{cory_experimental_1998}%
  \BibitemOpen
  \bibfield  {author} {\bibinfo {author} {\bibfnamefont {D.~G.}\ \bibnamefont
  {Cory}}, \bibinfo {author} {\bibfnamefont {M.~D.}\ \bibnamefont {Price}},
  \bibinfo {author} {\bibfnamefont {W.}~\bibnamefont {Maas}}, \bibinfo {author}
  {\bibfnamefont {E.}~\bibnamefont {Knill}}, \bibinfo {author} {\bibfnamefont
  {R.}~\bibnamefont {Laflamme}}, \bibinfo {author} {\bibfnamefont {W.~H.}\
  \bibnamefont {Zurek}}, \bibinfo {author} {\bibfnamefont {T.~F.}\ \bibnamefont
  {Havel}}, \ and\ \bibinfo {author} {\bibfnamefont {S.~S.}\ \bibnamefont
  {Somaroo}},\ }\href {\doibase 10.1103/PhysRevLett.81.2152} {\bibfield
  {journal} {\bibinfo  {journal} {Physical Review Letters}\ }\textbf {\bibinfo
  {volume} {81}},\ \bibinfo {pages} {2152} (\bibinfo {year}
  {1998})}\BibitemShut {NoStop}%
\bibitem [{\citenamefont {Chiaverini}\ \emph {et~al.}(2004)\citenamefont
  {Chiaverini}, \citenamefont {Leibfried}, \citenamefont {Schaetz},
  \citenamefont {Barrett}, \citenamefont {Blakestad}, \citenamefont {Britton},
  \citenamefont {Itano}, \citenamefont {Jost}, \citenamefont {Knill},
  \citenamefont {Langer}, \citenamefont {Ozeri},\ and\ \citenamefont
  {Wineland}}]{chiaverini_realization_2004}%
  \BibitemOpen
  \bibfield  {author} {\bibinfo {author} {\bibfnamefont {J.}~\bibnamefont
  {Chiaverini}}, \bibinfo {author} {\bibfnamefont {D.}~\bibnamefont
  {Leibfried}}, \bibinfo {author} {\bibfnamefont {T.}~\bibnamefont {Schaetz}},
  \bibinfo {author} {\bibfnamefont {M.~D.}\ \bibnamefont {Barrett}}, \bibinfo
  {author} {\bibfnamefont {R.~B.}\ \bibnamefont {Blakestad}}, \bibinfo {author}
  {\bibfnamefont {J.}~\bibnamefont {Britton}}, \bibinfo {author} {\bibfnamefont
  {W.~M.}\ \bibnamefont {Itano}}, \bibinfo {author} {\bibfnamefont {J.~D.}\
  \bibnamefont {Jost}}, \bibinfo {author} {\bibfnamefont {E.}~\bibnamefont
  {Knill}}, \bibinfo {author} {\bibfnamefont {C.}~\bibnamefont {Langer}},
  \bibinfo {author} {\bibfnamefont {R.}~\bibnamefont {Ozeri}}, \ and\ \bibinfo
  {author} {\bibfnamefont {D.~J.}\ \bibnamefont {Wineland}},\ }\href {\doibase
  10.1038/nature03074} {\bibfield  {journal} {\bibinfo  {journal} {Nature}\
  }\textbf {\bibinfo {volume} {432}},\ \bibinfo {pages} {602} (\bibinfo {year}
  {2004})}\BibitemShut {NoStop}%
\bibitem [{\citenamefont {Reed}\ \emph {et~al.}(2012)\citenamefont {Reed},
  \citenamefont {DiCarlo}, \citenamefont {Nigg}, \citenamefont {Sun},
  \citenamefont {Frunzio}, \citenamefont {Girvin},\ and\ \citenamefont
  {Schoelkopf}}]{reed_realization_2012}%
  \BibitemOpen
  \bibfield  {author} {\bibinfo {author} {\bibfnamefont {M.~D.}\ \bibnamefont
  {Reed}}, \bibinfo {author} {\bibfnamefont {L.}~\bibnamefont {DiCarlo}},
  \bibinfo {author} {\bibfnamefont {S.~E.}\ \bibnamefont {Nigg}}, \bibinfo
  {author} {\bibfnamefont {L.}~\bibnamefont {Sun}}, \bibinfo {author}
  {\bibfnamefont {L.}~\bibnamefont {Frunzio}}, \bibinfo {author} {\bibfnamefont
  {S.~M.}\ \bibnamefont {Girvin}}, \ and\ \bibinfo {author} {\bibfnamefont
  {R.~J.}\ \bibnamefont {Schoelkopf}},\ }\href {\doibase 10.1038/nature10786}
  {\bibfield  {journal} {\bibinfo  {journal} {Nature}\ }\textbf {\bibinfo
  {volume} {482}},\ \bibinfo {pages} {382} (\bibinfo {year}
  {2012})}\BibitemShut {NoStop}%
\bibitem [{\citenamefont {Barreiro}\ \emph {et~al.}(2011)\citenamefont
  {Barreiro}, \citenamefont {M{\"u}ller}, \citenamefont {Schindler},
  \citenamefont {Nigg}, \citenamefont {Monz}, \citenamefont {Chwalla},
  \citenamefont {Hennrich}, \citenamefont {Roos}, \citenamefont {Zoller},\ and\
  \citenamefont {Blatt}}]{barreiro_open-system_2011}%
  \BibitemOpen
  \bibfield  {author} {\bibinfo {author} {\bibfnamefont {J.~T.}\ \bibnamefont
  {Barreiro}}, \bibinfo {author} {\bibfnamefont {M.}~\bibnamefont
  {M{\"u}ller}}, \bibinfo {author} {\bibfnamefont {P.}~\bibnamefont
  {Schindler}}, \bibinfo {author} {\bibfnamefont {D.}~\bibnamefont {Nigg}},
  \bibinfo {author} {\bibfnamefont {T.}~\bibnamefont {Monz}}, \bibinfo {author}
  {\bibfnamefont {M.}~\bibnamefont {Chwalla}}, \bibinfo {author} {\bibfnamefont
  {M.}~\bibnamefont {Hennrich}}, \bibinfo {author} {\bibfnamefont {C.~F.}\
  \bibnamefont {Roos}}, \bibinfo {author} {\bibfnamefont {P.}~\bibnamefont
  {Zoller}}, \ and\ \bibinfo {author} {\bibfnamefont {R.}~\bibnamefont
  {Blatt}},\ }\href {\doibase 10.1038/nature09801} {\bibfield  {journal}
  {\bibinfo  {journal} {Nature}\ }\textbf {\bibinfo {volume} {470}},\ \bibinfo
  {pages} {486} (\bibinfo {year} {2011})}\BibitemShut {NoStop}%
\bibitem [{\citenamefont {Schindler}\ \emph {et~al.}(2011)\citenamefont
  {Schindler}, \citenamefont {Barreiro}, \citenamefont {Monz}, \citenamefont
  {Nebendahl}, \citenamefont {Nigg}, \citenamefont {Chwalla}, \citenamefont
  {Hennrich},\ and\ \citenamefont {Blatt}}]{schindler_experimental_2011}%
  \BibitemOpen
  \bibfield  {author} {\bibinfo {author} {\bibfnamefont {P.}~\bibnamefont
  {Schindler}}, \bibinfo {author} {\bibfnamefont {J.~T.}\ \bibnamefont
  {Barreiro}}, \bibinfo {author} {\bibfnamefont {T.}~\bibnamefont {Monz}},
  \bibinfo {author} {\bibfnamefont {V.}~\bibnamefont {Nebendahl}}, \bibinfo
  {author} {\bibfnamefont {D.}~\bibnamefont {Nigg}}, \bibinfo {author}
  {\bibfnamefont {M.}~\bibnamefont {Chwalla}}, \bibinfo {author} {\bibfnamefont
  {M.}~\bibnamefont {Hennrich}}, \ and\ \bibinfo {author} {\bibfnamefont
  {R.}~\bibnamefont {Blatt}},\ }\href {\doibase 10.1126/science.1203329}
  {\bibfield  {journal} {\bibinfo  {journal} {Science}\ }\textbf {\bibinfo
  {volume} {332}},\ \bibinfo {pages} {1059} (\bibinfo {year}
  {2011})}\BibitemShut {NoStop}%
\bibitem [{\citenamefont {Nigg}\ \emph {et~al.}(2014)\citenamefont {Nigg},
  \citenamefont {M{\"u}ller}, \citenamefont {Martinez}, \citenamefont
  {Schindler}, \citenamefont {Hennrich}, \citenamefont {Monz}, \citenamefont
  {Martin-Delgado},\ and\ \citenamefont {Blatt}}]{nigg_quantum_2014}%
  \BibitemOpen
  \bibfield  {author} {\bibinfo {author} {\bibfnamefont {D.}~\bibnamefont
  {Nigg}}, \bibinfo {author} {\bibfnamefont {M.}~\bibnamefont {M{\"u}ller}},
  \bibinfo {author} {\bibfnamefont {E.~A.}\ \bibnamefont {Martinez}}, \bibinfo
  {author} {\bibfnamefont {P.}~\bibnamefont {Schindler}}, \bibinfo {author}
  {\bibfnamefont {M.}~\bibnamefont {Hennrich}}, \bibinfo {author}
  {\bibfnamefont {T.}~\bibnamefont {Monz}}, \bibinfo {author} {\bibfnamefont
  {M.~A.}\ \bibnamefont {Martin-Delgado}}, \ and\ \bibinfo {author}
  {\bibfnamefont {R.}~\bibnamefont {Blatt}},\ }\href {\doibase
  10.1126/science.1253742} {\bibfield  {journal} {\bibinfo  {journal}
  {Science}\ }\textbf {\bibinfo {volume} {345}},\ \bibinfo {pages} {302}
  (\bibinfo {year} {2014})}\BibitemShut {NoStop}%
\bibitem [{\citenamefont {Pfaff}\ \emph {et~al.}(2013)\citenamefont {Pfaff},
  \citenamefont {Taminiau}, \citenamefont {Robledo}, \citenamefont {Bernien},
  \citenamefont {Markham}, \citenamefont {Twitchen},\ and\ \citenamefont
  {Hanson}}]{pfaff_demonstration_2013}%
  \BibitemOpen
  \bibfield  {author} {\bibinfo {author} {\bibfnamefont {W.}~\bibnamefont
  {Pfaff}}, \bibinfo {author} {\bibfnamefont {T.~H.}\ \bibnamefont {Taminiau}},
  \bibinfo {author} {\bibfnamefont {L.}~\bibnamefont {Robledo}}, \bibinfo
  {author} {\bibfnamefont {H.}~\bibnamefont {Bernien}}, \bibinfo {author}
  {\bibfnamefont {M.}~\bibnamefont {Markham}}, \bibinfo {author} {\bibfnamefont
  {D.~J.}\ \bibnamefont {Twitchen}}, \ and\ \bibinfo {author} {\bibfnamefont
  {R.}~\bibnamefont {Hanson}},\ }\href {\doibase 10.1038/nphys2444} {\bibfield
  {journal} {\bibinfo  {journal} {Nature Physics}\ }\textbf {\bibinfo {volume}
  {9}},\ \bibinfo {pages} {29} (\bibinfo {year} {2013})}\BibitemShut {NoStop}%
\bibitem [{\citenamefont {Rist{\`e}}\ \emph {et~al.}(2013)\citenamefont
  {Rist{\`e}}, \citenamefont {Dukalski}, \citenamefont {Watson}, \citenamefont
  {de~Lange}, \citenamefont {Tiggelman}, \citenamefont {Blanter}, \citenamefont
  {Lehnert}, \citenamefont {Schouten},\ and\ \citenamefont
  {DiCarlo}}]{riste_deterministic_2013}%
  \BibitemOpen
  \bibfield  {author} {\bibinfo {author} {\bibfnamefont {D.}~\bibnamefont
  {Rist{\`e}}}, \bibinfo {author} {\bibfnamefont {M.}~\bibnamefont {Dukalski}},
  \bibinfo {author} {\bibfnamefont {C.~A.}\ \bibnamefont {Watson}}, \bibinfo
  {author} {\bibfnamefont {G.}~\bibnamefont {de~Lange}}, \bibinfo {author}
  {\bibfnamefont {M.~J.}\ \bibnamefont {Tiggelman}}, \bibinfo {author}
  {\bibfnamefont {Y.~M.}\ \bibnamefont {Blanter}}, \bibinfo {author}
  {\bibfnamefont {K.~W.}\ \bibnamefont {Lehnert}}, \bibinfo {author}
  {\bibfnamefont {R.~N.}\ \bibnamefont {Schouten}}, \ and\ \bibinfo {author}
  {\bibfnamefont {L.}~\bibnamefont {DiCarlo}},\ }\href {\doibase
  10.1038/nature12513} {\bibfield  {journal} {\bibinfo  {journal} {Nature}\
  }\textbf {\bibinfo {volume} {502}},\ \bibinfo {pages} {350} (\bibinfo {year}
  {2013})}\BibitemShut {NoStop}%
\bibitem [{\citenamefont {Sun}\ \emph {et~al.}(2014)\citenamefont {Sun},
  \citenamefont {Petrenko}, \citenamefont {Leghtas}, \citenamefont {Vlastakis},
  \citenamefont {Kirchmair}, \citenamefont {Sliwa}, \citenamefont {Narla},
  \citenamefont {Hatridge}, \citenamefont {Shankar}, \citenamefont {Blumoff},
  \citenamefont {Frunzio}, \citenamefont {Mirrahimi}, \citenamefont {Devoret},\
  and\ \citenamefont {Schoelkopf}}]{sun_tracking_2014}%
  \BibitemOpen
  \bibfield  {author} {\bibinfo {author} {\bibfnamefont {L.}~\bibnamefont
  {Sun}}, \bibinfo {author} {\bibfnamefont {A.}~\bibnamefont {Petrenko}},
  \bibinfo {author} {\bibfnamefont {Z.}~\bibnamefont {Leghtas}}, \bibinfo
  {author} {\bibfnamefont {B.}~\bibnamefont {Vlastakis}}, \bibinfo {author}
  {\bibfnamefont {G.}~\bibnamefont {Kirchmair}}, \bibinfo {author}
  {\bibfnamefont {K.~M.}\ \bibnamefont {Sliwa}}, \bibinfo {author}
  {\bibfnamefont {A.}~\bibnamefont {Narla}}, \bibinfo {author} {\bibfnamefont
  {M.}~\bibnamefont {Hatridge}}, \bibinfo {author} {\bibfnamefont
  {S.}~\bibnamefont {Shankar}}, \bibinfo {author} {\bibfnamefont
  {J.}~\bibnamefont {Blumoff}}, \bibinfo {author} {\bibfnamefont
  {L.}~\bibnamefont {Frunzio}}, \bibinfo {author} {\bibfnamefont
  {M.}~\bibnamefont {Mirrahimi}}, \bibinfo {author} {\bibfnamefont {M.~H.}\
  \bibnamefont {Devoret}}, \ and\ \bibinfo {author} {\bibfnamefont {R.~J.}\
  \bibnamefont {Schoelkopf}},\ }\href {\doibase 10.1038/nature13436} {\bibfield
   {journal} {\bibinfo  {journal} {Nature}\ }\textbf {\bibinfo {volume}
  {511}},\ \bibinfo {pages} {444} (\bibinfo {year} {2014})}\BibitemShut
  {NoStop}%
\bibitem [{\citenamefont {Monz}\ \emph {et~al.}(2016)\citenamefont {Monz},
  \citenamefont {Nigg}, \citenamefont {Martinez}, \citenamefont {Brandl},
  \citenamefont {Schindler}, \citenamefont {Rines}, \citenamefont {Wang},
  \citenamefont {Chuang},\ and\ \citenamefont {Blatt}}]{monz_realization_2016}%
  \BibitemOpen
  \bibfield  {author} {\bibinfo {author} {\bibfnamefont {T.}~\bibnamefont
  {Monz}}, \bibinfo {author} {\bibfnamefont {D.}~\bibnamefont {Nigg}}, \bibinfo
  {author} {\bibfnamefont {E.~A.}\ \bibnamefont {Martinez}}, \bibinfo {author}
  {\bibfnamefont {M.~F.}\ \bibnamefont {Brandl}}, \bibinfo {author}
  {\bibfnamefont {P.}~\bibnamefont {Schindler}}, \bibinfo {author}
  {\bibfnamefont {R.}~\bibnamefont {Rines}}, \bibinfo {author} {\bibfnamefont
  {S.~X.}\ \bibnamefont {Wang}}, \bibinfo {author} {\bibfnamefont {I.~L.}\
  \bibnamefont {Chuang}}, \ and\ \bibinfo {author} {\bibfnamefont
  {R.}~\bibnamefont {Blatt}},\ }\href {\doibase 10.1126/science.aad9480}
  {\bibfield  {journal} {\bibinfo  {journal} {Science}\ }\textbf {\bibinfo
  {volume} {351}},\ \bibinfo {pages} {1068} (\bibinfo {year}
  {2016})}\BibitemShut {NoStop}%
\bibitem [{\citenamefont {Schmidt}\ \emph {et~al.}(2005)\citenamefont
  {Schmidt}, \citenamefont {Rosenband}, \citenamefont {Langer}, \citenamefont
  {Itano}, \citenamefont {Bergquist},\ and\ \citenamefont
  {Wineland}}]{schmidt_spectroscopy_2005}%
  \BibitemOpen
  \bibfield  {author} {\bibinfo {author} {\bibfnamefont {P.~O.}\ \bibnamefont
  {Schmidt}}, \bibinfo {author} {\bibfnamefont {T.}~\bibnamefont {Rosenband}},
  \bibinfo {author} {\bibfnamefont {C.}~\bibnamefont {Langer}}, \bibinfo
  {author} {\bibfnamefont {W.~M.}\ \bibnamefont {Itano}}, \bibinfo {author}
  {\bibfnamefont {J.~C.}\ \bibnamefont {Bergquist}}, \ and\ \bibinfo {author}
  {\bibfnamefont {D.~J.}\ \bibnamefont {Wineland}},\ }\href {\doibase
  10.1126/science.1114375} {\bibfield  {journal} {\bibinfo  {journal}
  {Science}\ }\textbf {\bibinfo {volume} {309}},\ \bibinfo {pages} {749}
  (\bibinfo {year} {2005})}\BibitemShut {NoStop}%
\bibitem [{\citenamefont {Ballance}\ \emph {et~al.}(2015)\citenamefont
  {Ballance}, \citenamefont {Sch{\"a}fer}, \citenamefont {Home}, \citenamefont
  {Szwer}, \citenamefont {Webster}, \citenamefont {Allcock}, \citenamefont
  {Linke}, \citenamefont {Harty}, \citenamefont {Aude~Craik}, \citenamefont
  {Stacey}, \citenamefont {Steane},\ and\ \citenamefont
  {Lucas}}]{ballance_hybrid_2015}%
  \BibitemOpen
  \bibfield  {author} {\bibinfo {author} {\bibfnamefont {C.~J.}\ \bibnamefont
  {Ballance}}, \bibinfo {author} {\bibfnamefont {V.~M.}\ \bibnamefont
  {Sch{\"a}fer}}, \bibinfo {author} {\bibfnamefont {J.~P.}\ \bibnamefont
  {Home}}, \bibinfo {author} {\bibfnamefont {D.~J.}\ \bibnamefont {Szwer}},
  \bibinfo {author} {\bibfnamefont {S.~C.}\ \bibnamefont {Webster}}, \bibinfo
  {author} {\bibfnamefont {D.~T.~C.}\ \bibnamefont {Allcock}}, \bibinfo
  {author} {\bibfnamefont {N.~M.}\ \bibnamefont {Linke}}, \bibinfo {author}
  {\bibfnamefont {T.~P.}\ \bibnamefont {Harty}}, \bibinfo {author}
  {\bibfnamefont {D.~P.~L.}\ \bibnamefont {Aude~Craik}}, \bibinfo {author}
  {\bibfnamefont {D.~N.}\ \bibnamefont {Stacey}}, \bibinfo {author}
  {\bibfnamefont {A.~M.}\ \bibnamefont {Steane}}, \ and\ \bibinfo {author}
  {\bibfnamefont {D.~M.}\ \bibnamefont {Lucas}},\ }\href {\doibase
  10.1038/nature16184} {\bibfield  {journal} {\bibinfo  {journal} {Nature}\
  }\textbf {\bibinfo {volume} {528}},\ \bibinfo {pages} {384} (\bibinfo {year}
  {2015})}\BibitemShut {NoStop}%
\bibitem [{\citenamefont {Gaebler}\ \emph {et~al.}(2016)\citenamefont
  {Gaebler}, \citenamefont {Tan}, \citenamefont {Lin}, \citenamefont {Wan},
  \citenamefont {Bowler}, \citenamefont {Keith}, \citenamefont {Glancy},
  \citenamefont {Coakley}, \citenamefont {Knill}, \citenamefont {Leibfried},\
  and\ \citenamefont {Wineland}}]{gaebler_high-fidelity_2016}%
  \BibitemOpen
  \bibfield  {author} {\bibinfo {author} {\bibfnamefont {J.~P.}\ \bibnamefont
  {Gaebler}}, \bibinfo {author} {\bibfnamefont {T.~R.}\ \bibnamefont {Tan}},
  \bibinfo {author} {\bibfnamefont {Y.}~\bibnamefont {Lin}}, \bibinfo {author}
  {\bibfnamefont {Y.}~\bibnamefont {Wan}}, \bibinfo {author} {\bibfnamefont
  {R.}~\bibnamefont {Bowler}}, \bibinfo {author} {\bibfnamefont {A.~C.}\
  \bibnamefont {Keith}}, \bibinfo {author} {\bibfnamefont {S.}~\bibnamefont
  {Glancy}}, \bibinfo {author} {\bibfnamefont {K.}~\bibnamefont {Coakley}},
  \bibinfo {author} {\bibfnamefont {E.}~\bibnamefont {Knill}}, \bibinfo
  {author} {\bibfnamefont {D.}~\bibnamefont {Leibfried}}, \ and\ \bibinfo
  {author} {\bibfnamefont {D.~J.}\ \bibnamefont {Wineland}},\ }\href {\doibase
  10.1103/PhysRevLett.117.060505} {\bibfield  {journal} {\bibinfo  {journal}
  {Physical Review Letters}\ }\textbf {\bibinfo {volume} {117}},\ \bibinfo
  {pages} {060505} (\bibinfo {year} {2016})}\BibitemShut {NoStop}%
\bibitem [{\citenamefont {Barrett}\ \emph {et~al.}(2003)\citenamefont
  {Barrett}, \citenamefont {DeMarco}, \citenamefont {Schaetz}, \citenamefont
  {Meyer}, \citenamefont {Leibfried}, \citenamefont {Britton}, \citenamefont
  {Chiaverini}, \citenamefont {Itano}, \citenamefont {Jelenkovi{\'c}},
  \citenamefont {Jost}, \citenamefont {Langer}, \citenamefont {Rosenband},\
  and\ \citenamefont {Wineland}}]{barrett_sympathetic_2003}%
  \BibitemOpen
  \bibfield  {author} {\bibinfo {author} {\bibfnamefont {M.~D.}\ \bibnamefont
  {Barrett}}, \bibinfo {author} {\bibfnamefont {B.}~\bibnamefont {DeMarco}},
  \bibinfo {author} {\bibfnamefont {T.}~\bibnamefont {Schaetz}}, \bibinfo
  {author} {\bibfnamefont {V.}~\bibnamefont {Meyer}}, \bibinfo {author}
  {\bibfnamefont {D.}~\bibnamefont {Leibfried}}, \bibinfo {author}
  {\bibfnamefont {J.}~\bibnamefont {Britton}}, \bibinfo {author} {\bibfnamefont
  {J.}~\bibnamefont {Chiaverini}}, \bibinfo {author} {\bibfnamefont {W.~M.}\
  \bibnamefont {Itano}}, \bibinfo {author} {\bibfnamefont {B.}~\bibnamefont
  {Jelenkovi{\'c}}}, \bibinfo {author} {\bibfnamefont {J.~D.}\ \bibnamefont
  {Jost}}, \bibinfo {author} {\bibfnamefont {C.}~\bibnamefont {Langer}},
  \bibinfo {author} {\bibfnamefont {T.}~\bibnamefont {Rosenband}}, \ and\
  \bibinfo {author} {\bibfnamefont {D.~J.}\ \bibnamefont {Wineland}},\ }\href
  {\doibase 10.1103/PhysRevA.68.042302} {\bibfield  {journal} {\bibinfo
  {journal} {Physical Review A}\ }\textbf {\bibinfo {volume} {68}},\ \bibinfo
  {pages} {042302} (\bibinfo {year} {2003})}\BibitemShut {NoStop}%
\bibitem [{\citenamefont {Home}\ \emph {et~al.}(2009)\citenamefont {Home},
  \citenamefont {Hanneke}, \citenamefont {Jost}, \citenamefont {Amini},
  \citenamefont {Leibfried},\ and\ \citenamefont
  {Wineland}}]{home_complete_2009}%
  \BibitemOpen
  \bibfield  {author} {\bibinfo {author} {\bibfnamefont {J.~P.}\ \bibnamefont
  {Home}}, \bibinfo {author} {\bibfnamefont {D.}~\bibnamefont {Hanneke}},
  \bibinfo {author} {\bibfnamefont {J.~D.}\ \bibnamefont {Jost}}, \bibinfo
  {author} {\bibfnamefont {J.~M.}\ \bibnamefont {Amini}}, \bibinfo {author}
  {\bibfnamefont {D.}~\bibnamefont {Leibfried}}, \ and\ \bibinfo {author}
  {\bibfnamefont {D.~J.}\ \bibnamefont {Wineland}},\ }\href {\doibase
  10.1126/science.1177077} {\bibfield  {journal} {\bibinfo  {journal}
  {Science}\ }\textbf {\bibinfo {volume} {325}},\ \bibinfo {pages} {1227}
  (\bibinfo {year} {2009})}\BibitemShut {NoStop}%
\bibitem [{\citenamefont {Hume}, \citenamefont {Rosenband},\ and\ \citenamefont
  {Wineland}(2007)}]{hume_high-fidelity_2007}%
  \BibitemOpen
  \bibfield  {author} {\bibinfo {author} {\bibfnamefont {D.~B.}\ \bibnamefont
  {Hume}}, \bibinfo {author} {\bibfnamefont {T.}~\bibnamefont {Rosenband}}, \
  and\ \bibinfo {author} {\bibfnamefont {D.~J.}\ \bibnamefont {Wineland}},\
  }\href {\doibase 10.1103/PhysRevLett.99.120502} {\bibfield  {journal}
  {\bibinfo  {journal} {Physical Review Letters}\ }\textbf {\bibinfo {volume}
  {99}},\ \bibinfo {pages} {120502} (\bibinfo {year} {2007})}\BibitemShut
  {NoStop}%
\bibitem [{\citenamefont {Langer}\ \emph {et~al.}(2005)\citenamefont {Langer},
  \citenamefont {Ozeri}, \citenamefont {Jost}, \citenamefont {Chiaverini},
  \citenamefont {DeMarco}, \citenamefont {Ben-Kish}, \citenamefont {Blakestad},
  \citenamefont {Britton}, \citenamefont {Hume}, \citenamefont {Itano},
  \citenamefont {Leibfried}, \citenamefont {Reichle}, \citenamefont
  {Rosenband}, \citenamefont {Schaetz}, \citenamefont {Schmidt},\ and\
  \citenamefont {Wineland}}]{langer_long-lived_2005}%
  \BibitemOpen
  \bibfield  {author} {\bibinfo {author} {\bibfnamefont {C.}~\bibnamefont
  {Langer}}, \bibinfo {author} {\bibfnamefont {R.}~\bibnamefont {Ozeri}},
  \bibinfo {author} {\bibfnamefont {J.~D.}\ \bibnamefont {Jost}}, \bibinfo
  {author} {\bibfnamefont {J.}~\bibnamefont {Chiaverini}}, \bibinfo {author}
  {\bibfnamefont {B.}~\bibnamefont {DeMarco}}, \bibinfo {author} {\bibfnamefont
  {A.}~\bibnamefont {Ben-Kish}}, \bibinfo {author} {\bibfnamefont {R.~B.}\
  \bibnamefont {Blakestad}}, \bibinfo {author} {\bibfnamefont {J.}~\bibnamefont
  {Britton}}, \bibinfo {author} {\bibfnamefont {D.~B.}\ \bibnamefont {Hume}},
  \bibinfo {author} {\bibfnamefont {W.~M.}\ \bibnamefont {Itano}}, \bibinfo
  {author} {\bibfnamefont {D.}~\bibnamefont {Leibfried}}, \bibinfo {author}
  {\bibfnamefont {R.}~\bibnamefont {Reichle}}, \bibinfo {author} {\bibfnamefont
  {T.}~\bibnamefont {Rosenband}}, \bibinfo {author} {\bibfnamefont
  {T.}~\bibnamefont {Schaetz}}, \bibinfo {author} {\bibfnamefont {P.~O.}\
  \bibnamefont {Schmidt}}, \ and\ \bibinfo {author} {\bibfnamefont {D.~J.}\
  \bibnamefont {Wineland}},\ }\href {\doibase 10.1103/PhysRevLett.95.060502}
  {\bibfield  {journal} {\bibinfo  {journal} {Physical Review Letters}\
  }\textbf {\bibinfo {volume} {95}},\ \bibinfo {pages} {060502} (\bibinfo
  {year} {2005})}\BibitemShut {NoStop}%
\bibitem [{\citenamefont {Bruzewicz}\ \emph {et~al.}(2017)\citenamefont
  {Bruzewicz}, \citenamefont {McConnell}, \citenamefont {Sedlacek},
  \citenamefont {Stuart}, \citenamefont {Loh}, \citenamefont {Sage},\ and\
  \citenamefont {Chiaverini}}]{bruzewicz_high-fidelity_2017}%
  \BibitemOpen
  \bibfield  {author} {\bibinfo {author} {\bibfnamefont {C.}~\bibnamefont
  {Bruzewicz}}, \bibinfo {author} {\bibfnamefont {R.}~\bibnamefont
  {McConnell}}, \bibinfo {author} {\bibfnamefont {J.}~\bibnamefont {Sedlacek}},
  \bibinfo {author} {\bibfnamefont {J.}~\bibnamefont {Stuart}}, \bibinfo
  {author} {\bibfnamefont {W.}~\bibnamefont {Loh}}, \bibinfo {author}
  {\bibfnamefont {J.}~\bibnamefont {Sage}}, \ and\ \bibinfo {author}
  {\bibfnamefont {J.}~\bibnamefont {Chiaverini}},\ }\href
  {http://arxiv.org/abs/1706.05102} {\bibfield  {journal} {\bibinfo  {journal}
  {arXiv:1706.05102 [physics, physics:quant-ph]}\ } (\bibinfo {year}
  {2017})}\BibitemShut {NoStop}%
\bibitem [{\citenamefont {Roos}\ \emph {et~al.}(2000)\citenamefont {Roos},
  \citenamefont {Leibfried}, \citenamefont {Mundt}, \citenamefont
  {Schmidt-Kaler}, \citenamefont {Eschner},\ and\ \citenamefont
  {Blatt}}]{roos_experimental_2000}%
  \BibitemOpen
  \bibfield  {author} {\bibinfo {author} {\bibfnamefont {C.~F.}\ \bibnamefont
  {Roos}}, \bibinfo {author} {\bibfnamefont {D.}~\bibnamefont {Leibfried}},
  \bibinfo {author} {\bibfnamefont {A.}~\bibnamefont {Mundt}}, \bibinfo
  {author} {\bibfnamefont {F.}~\bibnamefont {Schmidt-Kaler}}, \bibinfo {author}
  {\bibfnamefont {J.}~\bibnamefont {Eschner}}, \ and\ \bibinfo {author}
  {\bibfnamefont {R.}~\bibnamefont {Blatt}},\ }\href {\doibase
  10.1103/PhysRevLett.85.5547} {\bibfield  {journal} {\bibinfo  {journal}
  {Physical Review Letters}\ }\textbf {\bibinfo {volume} {85}},\ \bibinfo
  {pages} {5547} (\bibinfo {year} {2000})}\BibitemShut {NoStop}%
\bibitem [{\citenamefont {Tan}\ \emph {et~al.}(2015)\citenamefont {Tan},
  \citenamefont {Gaebler}, \citenamefont {Lin}, \citenamefont {Wan},
  \citenamefont {Bowler}, \citenamefont {Leibfried},\ and\ \citenamefont
  {Wineland}}]{tan_multi-element_2015}%
  \BibitemOpen
  \bibfield  {author} {\bibinfo {author} {\bibfnamefont {T.~R.}\ \bibnamefont
  {Tan}}, \bibinfo {author} {\bibfnamefont {J.~P.}\ \bibnamefont {Gaebler}},
  \bibinfo {author} {\bibfnamefont {Y.}~\bibnamefont {Lin}}, \bibinfo {author}
  {\bibfnamefont {Y.}~\bibnamefont {Wan}}, \bibinfo {author} {\bibfnamefont
  {R.}~\bibnamefont {Bowler}}, \bibinfo {author} {\bibfnamefont
  {D.}~\bibnamefont {Leibfried}}, \ and\ \bibinfo {author} {\bibfnamefont
  {D.~J.}\ \bibnamefont {Wineland}},\ }\href {\doibase 10.1038/nature16186}
  {\bibfield  {journal} {\bibinfo  {journal} {Nature}\ }\textbf {\bibinfo
  {volume} {528}},\ \bibinfo {pages} {380} (\bibinfo {year}
  {2015})}\BibitemShut {NoStop}%
\bibitem [{\citenamefont {S{\o}rensen}\ and\ \citenamefont
  {M{\o}lmer}(2000)}]{sorensen_entanglement_2000}%
  \BibitemOpen
  \bibfield  {author} {\bibinfo {author} {\bibfnamefont {A.}~\bibnamefont
  {S{\o}rensen}}\ and\ \bibinfo {author} {\bibfnamefont {K.}~\bibnamefont
  {M{\o}lmer}},\ }\href {\doibase 10.1103/PhysRevA.62.022311} {\bibfield
  {journal} {\bibinfo  {journal} {Physical Review A}\ }\textbf {\bibinfo
  {volume} {62}},\ \bibinfo {pages} {022311} (\bibinfo {year}
  {2000})}\BibitemShut {NoStop}%
\bibitem [{\citenamefont {Lee}\ \emph {et~al.}(2005)\citenamefont {Lee},
  \citenamefont {Brickman}, \citenamefont {Deslauriers}, \citenamefont
  {Haljan}, \citenamefont {Duan},\ and\ \citenamefont
  {Monroe}}]{lee_phase_2005}%
  \BibitemOpen
  \bibfield  {author} {\bibinfo {author} {\bibfnamefont {P.~J.}\ \bibnamefont
  {Lee}}, \bibinfo {author} {\bibfnamefont {K.-A.}\ \bibnamefont {Brickman}},
  \bibinfo {author} {\bibfnamefont {L.}~\bibnamefont {Deslauriers}}, \bibinfo
  {author} {\bibfnamefont {P.~C.}\ \bibnamefont {Haljan}}, \bibinfo {author}
  {\bibfnamefont {L.-M.}\ \bibnamefont {Duan}}, \ and\ \bibinfo {author}
  {\bibfnamefont {C.}~\bibnamefont {Monroe}},\ }\href {\doibase
  10.1088/1464-4266/7/10/025} {\bibfield  {journal} {\bibinfo  {journal}
  {Journal of Optics B: Quantum and Semiclassical Optics}\ }\textbf {\bibinfo
  {volume} {7}},\ \bibinfo {pages} {S371} (\bibinfo {year} {2005})}\BibitemShut
  {NoStop}%
\bibitem [{\citenamefont {Monroe}\ \emph {et~al.}(1995)\citenamefont {Monroe},
  \citenamefont {Meekhof}, \citenamefont {King}, \citenamefont {Jefferts},
  \citenamefont {Itano}, \citenamefont {Wineland},\ and\ \citenamefont
  {Gould}}]{monroe_resolved-sideband_1995}%
  \BibitemOpen
  \bibfield  {author} {\bibinfo {author} {\bibfnamefont {C.}~\bibnamefont
  {Monroe}}, \bibinfo {author} {\bibfnamefont {D.~M.}\ \bibnamefont {Meekhof}},
  \bibinfo {author} {\bibfnamefont {B.~E.}\ \bibnamefont {King}}, \bibinfo
  {author} {\bibfnamefont {S.~R.}\ \bibnamefont {Jefferts}}, \bibinfo {author}
  {\bibfnamefont {W.~M.}\ \bibnamefont {Itano}}, \bibinfo {author}
  {\bibfnamefont {D.~J.}\ \bibnamefont {Wineland}}, \ and\ \bibinfo {author}
  {\bibfnamefont {P.}~\bibnamefont {Gould}},\ }\href {\doibase
  10.1103/PhysRevLett.75.4011} {\bibfield  {journal} {\bibinfo  {journal}
  {Physical Review Letters}\ }\textbf {\bibinfo {volume} {75}},\ \bibinfo
  {pages} {4011} (\bibinfo {year} {1995})}\BibitemShut {NoStop}%
\bibitem [{\citenamefont {Ballance}\ \emph {et~al.}(2016)\citenamefont
  {Ballance}, \citenamefont {Harty}, \citenamefont {Linke}, \citenamefont
  {Sepiol},\ and\ \citenamefont {Lucas}}]{ballance_high-fidelity_2016}%
  \BibitemOpen
  \bibfield  {author} {\bibinfo {author} {\bibfnamefont {C.~J.}\ \bibnamefont
  {Ballance}}, \bibinfo {author} {\bibfnamefont {T.~P.}\ \bibnamefont {Harty}},
  \bibinfo {author} {\bibfnamefont {N.~M.}\ \bibnamefont {Linke}}, \bibinfo
  {author} {\bibfnamefont {M.~A.}\ \bibnamefont {Sepiol}}, \ and\ \bibinfo
  {author} {\bibfnamefont {D.~M.}\ \bibnamefont {Lucas}},\ }\href {\doibase
  10.1103/PhysRevLett.117.060504} {\bibfield  {journal} {\bibinfo  {journal}
  {Physical Review Letters}\ }\textbf {\bibinfo {volume} {117}},\ \bibinfo
  {pages} {060504} (\bibinfo {year} {2016})}\BibitemShut {NoStop}%
\bibitem [{\citenamefont {Brown}\ and\ \citenamefont
  {Brown}(2018)}]{brown_comparing_2018}%
  \BibitemOpen
  \bibfield  {author} {\bibinfo {author} {\bibfnamefont {N.~C.}\ \bibnamefont
  {Brown}}\ and\ \bibinfo {author} {\bibfnamefont {K.~R.}\ \bibnamefont
  {Brown}},\ }\href {http://arxiv.org/abs/1803.02545} {\bibfield  {journal}
  {\bibinfo  {journal} {arXiv:1803.02545 [quant-ph]}\ } (\bibinfo {year}
  {2018})}\BibitemShut {NoStop}%
\bibitem [{\citenamefont {Bombin}\ and\ \citenamefont
  {Martin-Delgado}(2006)}]{bombin_topological_2006}%
  \BibitemOpen
  \bibfield  {author} {\bibinfo {author} {\bibfnamefont {H.}~\bibnamefont
  {Bombin}}\ and\ \bibinfo {author} {\bibfnamefont {M.~A.}\ \bibnamefont
  {Martin-Delgado}},\ }\href {\doibase 10.1103/PhysRevLett.97.180501}
  {\bibfield  {journal} {\bibinfo  {journal} {Physical Review Letters}\
  }\textbf {\bibinfo {volume} {97}},\ \bibinfo {pages} {180501} (\bibinfo
  {year} {2006})}\BibitemShut {NoStop}%
\bibitem [{\citenamefont {Fowler}\ \emph {et~al.}(2012)\citenamefont {Fowler},
  \citenamefont {Mariantoni}, \citenamefont {Martinis},\ and\ \citenamefont
  {Cleland}}]{fowler_surface_2012}%
  \BibitemOpen
  \bibfield  {author} {\bibinfo {author} {\bibfnamefont {A.~G.}\ \bibnamefont
  {Fowler}}, \bibinfo {author} {\bibfnamefont {M.}~\bibnamefont {Mariantoni}},
  \bibinfo {author} {\bibfnamefont {J.~M.}\ \bibnamefont {Martinis}}, \ and\
  \bibinfo {author} {\bibfnamefont {A.~N.}\ \bibnamefont {Cleland}},\ }\href
  {\doibase 10.1103/PhysRevA.86.032324} {\bibfield  {journal} {\bibinfo
  {journal} {Physical Review A}\ }\textbf {\bibinfo {volume} {86}},\ \bibinfo
  {pages} {032324} (\bibinfo {year} {2012})}\BibitemShut {NoStop}%
\bibitem [{\citenamefont {Raussendorf}\ and\ \citenamefont
  {Briegel}(2001)}]{raussendorf_one-way_2001}%
  \BibitemOpen
  \bibfield  {author} {\bibinfo {author} {\bibfnamefont {R.}~\bibnamefont
  {Raussendorf}}\ and\ \bibinfo {author} {\bibfnamefont {H.~J.}\ \bibnamefont
  {Briegel}},\ }\href {\doibase 10.1103/PhysRevLett.86.5188} {\bibfield
  {journal} {\bibinfo  {journal} {Physical Review Letters}\ }\textbf {\bibinfo
  {volume} {86}},\ \bibinfo {pages} {5188} (\bibinfo {year}
  {2001})}\BibitemShut {NoStop}%
\bibitem [{\citenamefont {Lanyon}\ \emph {et~al.}(2013)\citenamefont {Lanyon},
  \citenamefont {Jurcevic}, \citenamefont {Zwerger}, \citenamefont {Hempel},
  \citenamefont {Martinez}, \citenamefont {D{\"u}r}, \citenamefont {Briegel},
  \citenamefont {Blatt},\ and\ \citenamefont
  {Roos}}]{lanyon_measurement-based_2013}%
  \BibitemOpen
  \bibfield  {author} {\bibinfo {author} {\bibfnamefont {B.~P.}\ \bibnamefont
  {Lanyon}}, \bibinfo {author} {\bibfnamefont {P.}~\bibnamefont {Jurcevic}},
  \bibinfo {author} {\bibfnamefont {M.}~\bibnamefont {Zwerger}}, \bibinfo
  {author} {\bibfnamefont {C.}~\bibnamefont {Hempel}}, \bibinfo {author}
  {\bibfnamefont {E.~A.}\ \bibnamefont {Martinez}}, \bibinfo {author}
  {\bibfnamefont {W.}~\bibnamefont {D{\"u}r}}, \bibinfo {author} {\bibfnamefont
  {H.~J.}\ \bibnamefont {Briegel}}, \bibinfo {author} {\bibfnamefont
  {R.}~\bibnamefont {Blatt}}, \ and\ \bibinfo {author} {\bibfnamefont {C.~F.}\
  \bibnamefont {Roos}},\ }\href {\doibase 10.1103/PhysRevLett.111.210501}
  {\bibfield  {journal} {\bibinfo  {journal} {Physical Review Letters}\
  }\textbf {\bibinfo {volume} {111}},\ \bibinfo {pages} {210501} (\bibinfo
  {year} {2013})}\BibitemShut {NoStop}%
\bibitem [{\citenamefont {Gottesman}\ and\ \citenamefont
  {Chuang}(1999)}]{gottesman_demonstrating_1999}%
  \BibitemOpen
  \bibfield  {author} {\bibinfo {author} {\bibfnamefont {D.}~\bibnamefont
  {Gottesman}}\ and\ \bibinfo {author} {\bibfnamefont {I.~L.}\ \bibnamefont
  {Chuang}},\ }\href {\doibase 10.1038/46503} {\bibfield  {journal} {\bibinfo
  {journal} {Nature}\ }\textbf {\bibinfo {volume} {402}},\ \bibinfo {pages}
  {390} (\bibinfo {year} {1999})}\BibitemShut {NoStop}%
\bibitem [{\citenamefont {de~Clercq}\ \emph {et~al.}(2016)\citenamefont
  {de~Clercq}, \citenamefont {Lo}, \citenamefont {Marinelli}, \citenamefont
  {Nadlinger}, \citenamefont {Oswald}, \citenamefont {Negnevitsky},
  \citenamefont {Kienzler}, \citenamefont {Keitch},\ and\ \citenamefont
  {Home}}]{de_clercq_parallel_2016}%
  \BibitemOpen
  \bibfield  {author} {\bibinfo {author} {\bibfnamefont {L.~E.}\ \bibnamefont
  {de~Clercq}}, \bibinfo {author} {\bibfnamefont {H.-Y.}\ \bibnamefont {Lo}},
  \bibinfo {author} {\bibfnamefont {M.}~\bibnamefont {Marinelli}}, \bibinfo
  {author} {\bibfnamefont {D.}~\bibnamefont {Nadlinger}}, \bibinfo {author}
  {\bibfnamefont {R.}~\bibnamefont {Oswald}}, \bibinfo {author} {\bibfnamefont
  {V.}~\bibnamefont {Negnevitsky}}, \bibinfo {author} {\bibfnamefont
  {D.}~\bibnamefont {Kienzler}}, \bibinfo {author} {\bibfnamefont
  {B.}~\bibnamefont {Keitch}}, \ and\ \bibinfo {author} {\bibfnamefont {J.~P.}\
  \bibnamefont {Home}},\ }\href {\doibase 10.1103/PhysRevLett.116.080502}
  {\bibfield  {journal} {\bibinfo  {journal} {Physical Review Letters}\
  }\textbf {\bibinfo {volume} {116}},\ \bibinfo {pages} {080502} (\bibinfo
  {year} {2016})}\BibitemShut {NoStop}%
\bibitem [{\citenamefont {Home}(2013)}]{home_chapter_2013}%
  \BibitemOpen
  \bibfield  {author} {\bibinfo {author} {\bibfnamefont {J.~P.}\ \bibnamefont
  {Home}},\ }in\ \href
  {http://www.sciencedirect.com/science/article/pii/B9780124080904000049}
  {\emph {\bibinfo {booktitle} {Advances {In} {Atomic}, {Molecular}, and
  {Optical} {Physics}}}},\ \bibinfo {series} {Advances in {Atomic},
  {Molecular}, and {Optical} {Physics}}, Vol.\ \bibinfo {volume} {Volume 62},\
  \bibinfo {editor} {edited by\ \bibinfo {editor} {\bibfnamefont {P.~R. B. a.
  C. C.~L.}\ \bibnamefont {Ennio~Arimondo}}}\ (\bibinfo  {publisher} {Academic
  Press},\ \bibinfo {year} {2013})\ pp.\ \bibinfo {pages}
  {231--277}\BibitemShut {NoStop}%
\bibitem [{\citenamefont {Kienzler}\ \emph {et~al.}(2015)\citenamefont
  {Kienzler}, \citenamefont {Lo}, \citenamefont {Keitch}, \citenamefont
  {de~Clercq}, \citenamefont {Leupold}, \citenamefont {Lindenfelser},
  \citenamefont {Marinelli}, \citenamefont {Negnevitsky},\ and\ \citenamefont
  {Home}}]{kienzler_quantum_2015}%
  \BibitemOpen
  \bibfield  {author} {\bibinfo {author} {\bibfnamefont {D.}~\bibnamefont
  {Kienzler}}, \bibinfo {author} {\bibfnamefont {H.-Y.}\ \bibnamefont {Lo}},
  \bibinfo {author} {\bibfnamefont {B.}~\bibnamefont {Keitch}}, \bibinfo
  {author} {\bibfnamefont {L.~E.}\ \bibnamefont {de~Clercq}}, \bibinfo {author}
  {\bibfnamefont {F.}~\bibnamefont {Leupold}}, \bibinfo {author} {\bibfnamefont
  {F.}~\bibnamefont {Lindenfelser}}, \bibinfo {author} {\bibfnamefont
  {M.}~\bibnamefont {Marinelli}}, \bibinfo {author} {\bibfnamefont
  {V.}~\bibnamefont {Negnevitsky}}, \ and\ \bibinfo {author} {\bibfnamefont
  {J.~P.}\ \bibnamefont {Home}},\ }\href {\doibase 10.1126/science.1261033}
  {\bibfield  {journal} {\bibinfo  {journal} {Science}\ }\textbf {\bibinfo
  {volume} {347}},\ \bibinfo {pages} {53} (\bibinfo {year} {2015})}\BibitemShut
  {NoStop}%
\bibitem [{\citenamefont {Keitch}, \citenamefont {Negnevitsky},\ and\
  \citenamefont {Zhang}(2017)}]{keitch_programmable_2017}%
  \BibitemOpen
  \bibfield  {author} {\bibinfo {author} {\bibfnamefont {B.}~\bibnamefont
  {Keitch}}, \bibinfo {author} {\bibfnamefont {V.}~\bibnamefont {Negnevitsky}},
  \ and\ \bibinfo {author} {\bibfnamefont {W.}~\bibnamefont {Zhang}},\ }\href
  {http://arxiv.org/abs/1710.04282} {\bibfield  {journal} {\bibinfo  {journal}
  {arXiv:1710.04282 [physics, physics:quant-ph]}\ } (\bibinfo {year}
  {2017})}\BibitemShut {NoStop}%
\bibitem [{\citenamefont {Ralph}\ \emph {et~al.}(2006)\citenamefont {Ralph},
  \citenamefont {Bartlett}, \citenamefont {O{\textquoteright}Brien},
  \citenamefont {Pryde},\ and\ \citenamefont {Wiseman}}]{ralph_quantum_2006}%
  \BibitemOpen
  \bibfield  {author} {\bibinfo {author} {\bibfnamefont {T.~C.}\ \bibnamefont
  {Ralph}}, \bibinfo {author} {\bibfnamefont {S.~D.}\ \bibnamefont {Bartlett}},
  \bibinfo {author} {\bibfnamefont {J.~L.}\ \bibnamefont
  {O{\textquoteright}Brien}}, \bibinfo {author} {\bibfnamefont {G.~J.}\
  \bibnamefont {Pryde}}, \ and\ \bibinfo {author} {\bibfnamefont {H.~M.}\
  \bibnamefont {Wiseman}},\ }\href {\doibase 10.1103/PhysRevA.73.012113}
  {\bibfield  {journal} {\bibinfo  {journal} {Physical Review A}\ }\textbf
  {\bibinfo {volume} {73}},\ \bibinfo {pages} {012113} (\bibinfo {year}
  {2006})}\BibitemShut {NoStop}%
\bibitem [{\citenamefont {Nielsen}\ and\ \citenamefont
  {Chuang}(2000)}]{nielsen_quantum_2000}%
  \BibitemOpen
  \bibfield  {author} {\bibinfo {author} {\bibfnamefont {M.~A.}\ \bibnamefont
  {Nielsen}}\ and\ \bibinfo {author} {\bibfnamefont {I.~L.}\ \bibnamefont
  {Chuang}},\ }\href@noop {} {\emph {\bibinfo {title} {Quantum {Computation}
  and {Quantum} {Information}}}}\ (\bibinfo  {publisher} {Cambridge University
  Press},\ \bibinfo {year} {2000})\BibitemShut {NoStop}%
\end{thebibliography}%

\end{document}